\documentclass[lettersize,journal]{IEEEtran}
\usepackage{amsmath,amsfonts}
\usepackage{algorithm}
\usepackage{amsthm}
\usepackage{algpseudocode}
\usepackage{array}
\usepackage[caption=false,font=normalsize,labelfont=sf,textfont=sf]{subfig}
\usepackage{textcomp}
\usepackage{stfloats}
\usepackage[dvipsnames]{xcolor}

\usepackage{cite}
\usepackage{url}

\usepackage{multirow}
\usepackage{verbatim}
\usepackage{graphicx}

\def\BibTeX{{\rm B\kern-.05em{\sc i\kern-.025em b}\kern-.08em
    T\kern-.1667em\lower.7ex\hbox{E}\kern-.125emX}}
\usepackage{balance}
\usepackage[normalem]{ulem}
\graphicspath{{fig/}}
\newtheorem{definition}{Definition}

\newtheorem{remark}{Remark}
\begin{document}
\title{Communication Strategy on Macro-and-Micro Traffic State in Cooperative Deep Reinforcement Learning for Regional Traffic Signal Control}
\author{Hankang Gu, Shangbo Wang ~\IEEEmembership{Member,~IEEE}, Dongyao Jia,~\IEEEmembership{Member,~IEEE}, Yuli Zhang, Yanrong Luo, \\Guoqiang Mao, ~\IEEEmembership{Fellow,~IEEE}, Jianping Wang,~\IEEEmembership{Fellow,~IEEE}, Eng Gee Lim, ~\IEEEmembership{Senior Member,~IEEE}

\thanks{Hankang Gu is with the School of Advanced Technology, Xi'an Jiaotong-Liverpool University, Suzhou 215123, China, also with the Department of Computer Science, University of Liverpool,
L69 3GJ Liverpool, U.K.(email: Hankang.Gu16@student.xjtlu.edu.cn)}
\thanks{Shangbo Wang is with the
Department of Engineering and Design, University of Sussex, BN1 9RH
Brighton, U.K. (e-mail: shangbo.wang@sussex.ac.uk)}
\thanks{Dongyao Jia and Yuli Zhang are with the School of Advanced Technology, Yanrong Luo is with International Business School Suzhou, Xi'an Jiaotong-Liverpool University, Suzhou 215123, China, (e-mail: Dongyao.Jia@xjtlu.edu.cn; \{Yuli.Zhang20, Yanrong.Luo21\}@student.xjtlu.edu.cn)}
\thanks{Guoqiang Mao is with the ISN State Key Lab, Xidian University, Xi'an 710126, China (email:gqmao@xidian.edu.cn)}
\thanks{Jianping~Wang is with the Department of Computer Science, City University of Hong Kong, Kowloon, Hong Kong SAR, China, and City University of Hong Kong Shenzhen Research Institute. (e-mail:  jianwang@cityu.edu.hk)}
}
\maketitle
\begin{abstract}
Adaptive Traffic Signal Control (ATSC) has become a popular research topic in intelligent transportation systems.
Regional Traffic Signal Control (RTSC) using the Multi-agent Deep Reinforcement Learning (MADRL) technique has become a promising approach for ATSC due to its ability to achieve the optimum trade-off between scalability and optimality. 
Most existing RTSC approaches partition a traffic network into several disjoint regions, followed by applying centralized reinforcement learning techniques to each region.
However, the pursuit of cooperation among RTSC agents still remains an open issue and no communication strategy for RTSC agents has been investigated.
In this paper, we propose communication strategies to capture the correlation of micro-traffic states among lanes and the correlation of macro-traffic states among intersections. We first justify the evolution equation of the RTSC process is Markovian via a system of store-and-forward queues. Next, based on the evolution equation, we propose two GAT-Aggregated (GA2) communication modules---GA2-Naive and GA2-Aug
to extract both intra-region and inter-region correlations between macro and micro traffic states. While GA2-Naive only considers the movements at each intersection, GA2-Aug also considers the lane-changing behavior of vehicles. Two proposed communication modules are then aggregated into two existing novel RTSC frameworks---RegionLight and Regional-DRL. Experimental results demonstrate that both GA2-Naive and GA2-Aug effectively improve the performance of existing RTSC frameworks under both real and synthetic scenarios. Hyperparameter testing also reveals the robustness and potential of our communication modules in large-scale traffic networks.

\end{abstract}
\section{Introduction}
\IEEEPARstart{W}ITH rapid urbanization and population growth in recent years, traffic congestion is becoming a prominent issue agitating all participants in the transportation system\cite{hussain2023investigating, smog}. 
The alleviation of traffic congestion brings both economic and environmental benefits\cite{geng2016environmental,inrix2021uk,brilon2013experiences}. Motivated by the urgent need, intelligent transportation systems have been widely studied to improve transportation efficiency by exploring optimal traffic flow control and optimal traffic signal control (TSC). TSC is a promising and cost-efficient approach where the vehicles' movements at each intersection are managed by traffic signals\cite{haydari2020deep}.  
Conventional TSC techniques such as GreenWave\cite{roess2004traffic} and Maxband \cite{little1981maxband} focus on rule-based control strategies which usually cast predefined assumptions on expected travel speeds or traffic cycle lengths\cite{wei2019survey}. However, traffic dynamics in real scenarios are much more complex. Consequently, conventional TSC techniques have limitations in adapting to these complicated conditions\cite{abdoos2011traffic}.

The recent rapid development of model-free deep reinforcement learning (DRL) techniques, which can adapt to large high dimensional states, has demonstrated significant potential in various research areas, including autonomous driving \cite{kiran2021deep} and cyber security \cite{nguyen2021deep}. 
The agent of the DRL technique makes sequential decisions in the Markov decision process (MDP) through a trial-and-error procedure\cite{sutton2018reinforcement,mnih2015human}. Single-agent reinforcement learning (RL) techniques have been applied to scenarios involving either one isolated intersection or several connected intersections\cite{li2016traffic,genders2016using,mousavi2017traffic}.
These completely centralized RL techniques exhibit a good convergence rate in small-scale traffic networks. However, as the scale of the traffic networks increases, the growth of traffic state space and joint action space becomes exponential, making the search for a joint optimal policy for all signals computationally impractical\cite{haydari2020deep}. 

To alleviate the scalability issue of completely centralized RL techniques, multi-agent deep reinforcement learning (MADRL) techniques have been proposed and studied by numerous researchers\cite{wei2019survey,haydari2020deep,li2023adaptive,wei2018intellilight,zhang2022neighborhood,ge2021multi}.  
Most existing MADRL techniques apply completely decentralized strategies in which one agent is assigned to control one specific intersection. The optimal joint action for the entire traffic network is the union of the optimal actions for each agent. Although the scalability issue in MADRL is alleviated, the environment becomes non-stationary due to the intricate interactions between agents\cite{gronauer2022multi}. Independent RL (IRL) agents even face theoretical convergence failure because each agent maximizes only its own rewards without considering the impact on other agents\cite{tan1993multi}. To foster cooperation among agents, various communication and coordination strategies are examined. This involves either the exchange of information between agents or the pursuit of an optimal joint action facilitated by coordinators\cite{oroojlooy2023review,zhu2024survey,zhang2023large}. Nonetheless, certain cooperative agents fail to converge when the number of agents becomes substantial\cite{chen2020toward}.

To balance scalability and optimality, regional traffic signal control (RTSC) is a compromised method typically involving two stages\cite{chu2016large,tan2019cooperative,gu2024large}.
The first stage partitions a large network into several disjoint smaller regions, each comprising a set of intersections. One straightforward way is to partition the traffic network into several fixed-shape regions\cite{tan2019cooperative}. However, fixed-shape regions lack adaptability. The regions in \cite{chu2016large} are partitioned by grouping intersections with internal strong traffic density dynamically and the regions in \cite{gu2024large} are partitioned by only constraining the topology of each region. 
After the network is partitioned, a centralized DRL technique is applied to control each region. 
While Regional-DRL (RDRL)\cite{tan2019cooperative} and RegionLight\cite{gu2024large} continue to utilize decentralized independent agents, cooperative deep reinforcement learning framework (Coder) implements a decentralized-to-centralized coordinator to estimate the global Q-value for the entire traffic network\cite{tan2019cooperative}. The regional control methods have successfully converged and identified globally optimal actions in large-scale traffic networks, managing up to 24 intersections \cite{tan2019cooperative} and up to 48 intersections \cite{gu2024large}. 
However, current regional control methods still exhibit the following limitations:

\begin{itemize}
    \item The model-free DRL agent interacts with the environment through trial-and-error procedures. Thus, modeling the nature of the environment as one MDP is crucial for the agent's convergence and performance. The assumption of MDP for an isolated intersection was justified in \cite{wei2019presslight} using a store-and-forward model \cite{varaiya2013max}. If the agent is assigned to control a region of signals, it has to consider the interactions among signals either inside the region (intra-region) or outside the regions (inter-region). However, the characterization of RTSC as the MDP has not yet been formally justified. 
    \item The development of cooperative regional signal control agents still faces a great challenge. One global coordinator is utilized to develop cooperative regional signal control agents by estimating the global Q-value\cite{tan2019cooperative}. However, searching for the optimal joint global action necessitates multiple rounds of estimations on different combinations of regional sub-optimal actions, and the convergence of the global coordinator is not yet guaranteed. Unlike coordination strategies, communication strategies enable agents to exchange specific information, thereby alleviating non-stationarity. However, no communication strategy between RTSC agents has yet been studied. 
\end{itemize}
To enhance cooperation among regional control agents, we first justify that the signal regional control process can be modeled as a Markov chain through a system of store-and-forward queueing models. Then, based on the evolution pattern of the Markov chain, we utilize the graph attention layer(GAT) to capture the correlations between different regions, considering both macro and micro traffic states. More specifically, our main contributions are listed as follows:
\begin{enumerate}
    \item
    In traffic networks, vehicles transit between lanes, moving from one incoming lane to one designated outgoing lane at each intersection. Once in an outgoing lane, vehicles can shift to any adjacent lanes. To characterize these traffic flow dynamics, we define both the movement matrix and the routing proportion matrix. Subsequently, we formulate the updating equation for the signal regional control process using a system of store-and-forward queuing models, and we demonstrate that the updating equation exhibits the Markov property.
    
    \item Based on the updating equation of the RTSC process, we further propose two novel communication modules ---GA2-Naive and GA2-Augmented (GA2-Aug) that capture the correlations of lane-level micro-traffic states and those of intersection-level macro-traffic states. 
    The micro-traffic state is the number of vehicles within each lane segmentation. 
    The macro-traffic state is the number of moving and waiting vehicles on lanes.  
    Then, we utilize GAT to aggregate micro and macro traffic states. More specifically, in Naive-GA2, the micro-traffic state is aggregated by involving the movement matrix, and the macro-traffic state is aggregated by involving adjacency between intersections. Additionally, the lane-changing behavior of vehicles is involved by using adjacency between lanes in Augmented-GA2.
    Finally, we aggregate two proposed communication modules with RegionLight and R-DRL frameworks.
    
    \item We evaluate our model on $4\times4$ and $16\times3$ grid traffic networks with both real and synthetic traffic flows. Empirical results show that the proposed communication modules improve the performance of RTSC models. We further examine the stability of our model by using different hyper-parameter settings,  typically on the number of multi-attention heads in the GAT and the number of cells in lane segmentations.
\end{enumerate}
The rest of the paper is organized as follows: Section \ref{sec: related work} reviews the related work on communication and coordination strategies for MADRL-based TSC models. Section \ref{sec: background} introduces the background of TSC and MADRL. Section \ref{sec: regional markovian property} presents the formal justification of the Markov decision process (MDP) in the regional signal control process. Section \ref{sec: communication strategy} describes the communication strategy developed based on the evolution equation formulated in the previous section. Section \ref{sec: experiment and result} outlines the experimental setup and discusses our findings. Finally, Section \ref{sec: conclusion} summarizes the paper.

\section{Related Work}
\label{sec: related work}
In this section, we mainly review the related work on cooperative MADRL-based TSC models that apply either communication or coordination strategies. We exclude IRL methods because they encounter convergence problems due to the non-stationary issue.

\subsection{Communication Strategy}
We review the related work on communication strategies, focusing primarily on three aspects: who to exchange with, what information to exchange, and how to exchange it.

One straightforward approach is to augment the observation of one agent by concatenating it with the observations of its neighboring agents\cite{arel2010reinforcement}. In contrast, some studies discriminate the contributions of neighboring agents by augmenting observations with weighted values.
Zhang et al. extended Hysteretic DQN (HDQN)\cite{omidshafiei2017deep} to neighborhood cooperative hysteretic DQN (NC-HDQN) by considering the correlation between two neighboring intersections\cite{zhang2022neighborhood}. In their work, the observation of one intersection is concatenated with the observation of its neighboring intersections, weighted by correlation degree. They further proposed a rule-based method, namely empirical NC-HDQN (ENC-HDQN), and a Pearson-correlation-coefficient-based method, namely Pearson NC-HDQN (PNC-HDQN). In ENC-HDQN, the correlation degree is defined based on the number of waiting vehicles between two intersections with a pre-defined threshold.  In contrast, PHC-HDQN collects the short-term reward trajectories for each agent and then applies the Pearson method to compute the correlations between neighboring intersections.  

Instead of concatenating neighboring information directly to local observation, the following studies encode neighboring information through neural networks.  
CoLight utilizes a stack of GAT to embed the observation of each agent by incorporating a dynamically weighted average of the observations from its neighboring agents\cite{wei2019colight}. 
Zhou et al. proposed Multi-agent Incentive Communication Deep Reinforcement Learning (MICDRL) to enable agents to create customized messages\cite{zhou2024cooperative}. MICDRL utilizes a multivariate Gaussian distribution (MGD) to infer other agents' actions based on their local information. The local Q-value is then combined with the weighted messages from neighboring agents, which are computed using the MGD.
Similarly, Mess-Net was proposed in Information Exchange Deep Q-Network (IEDQN) to facilitate information exchange among all agents \cite{xie2020iedqn}. In this approach, the current timestep observation and the previous timestep Q-value for each agent are first concatenated and embedded as local information. Then, the local information from all agents is concatenated and embedded centrally as a message block. This message block is subsequently divided into several message vectors, evenly allocated to all agents. Finally, each agent predicts its Q-value based on its observation and the corresponding message vector.

To further enhance communication, the following studies further exchange local policies or historical information.
In \cite{wang2021traffic}, the actor-critic agent considers its neighboring agents' observations and their policies.
Spatialtemporal correlations between agents are considered in NeurComm\cite{chu2020multi}. At each time step, the observations, historical hidden states, and previous timestep policies of the agent and its neighboring agents are merged and embedded as current hidden states. The spatiotemporal hidden state is then used to predict the state value.
Zhang et al. proposed the off-policy Nash deep Q-network (OPNDQN) which utilizes a fictitious play approach to increase the local agent's rewards without reducing those of its neighborhood\cite{zhang2023large}. The agents in OPNDQN exchange actions and OPNDQN also facilitates reaching a Nash equilibrium.
The agents in \cite{li2023adaptive} exchange information with their neighboring agents by determining the corresponding distances and utilizing mix-encoders to aggregate messages. 

\subsection{Coordination Strategy}
Apart from communication strategies, many researchers have studied the nature of the interactions between agents and proposed various coordination strategies to choose global joint action. Some studies assume the global Q-value of joint action is the sum of the Q-value of each local action.
The max-plus algorithm and transfer planning are applied to optimize the joint global action based on factorized global Q-value\cite{van2016coordinated}.
Lee et al. proposed a more straightforward method for computing the global Q-value \cite{lee2019reinforcement}. In their approach, the Q-values of all possible joint actions are first calculated by summing all local Q-values. The optimal joint action is then identified as the one with the highest global Q-value.

Another common strategy is to utilize one parameterized global coordinator to evaluate the global Q-value for global joint action, allowing for more flexible assumptions.
Li et al. proposed an Adaptive Multi-agent Deep Mixed Reinforcement Learning (AMDMRL) model using a mixed state-action value function inspired by QMIX\cite{rashid2020monotonic}\cite{li2023adaptive}. The mixed state-action value assumes all agents contribute positively to the global Q-value, implying that there is no competition between these agents.
Cooperative deep reinforcement learning (Coder) is proposed to take the last hidden layers of all agents and predict the global Q-value without the above assumptions \cite{tan2019cooperative}. Meanwhile, the Coder initially collects several local sub-optimal actions proposed by agents and then estimates the global Q-values of different combinations of these proposed actions through an Iterative Action Search process.

\section{Preliminaries}
\label{sec: background}

\subsection{TSC}
\label{sec: Traffic Signal Control Definition}
A traffic network is defined as a directed graph $\mathcal{G}=(\mathcal{V},\mathcal{E})$ where $v \in \mathcal{V}$ represents an intersection and $e_{vu}=(v,u)\in  \mathcal{E}$ represents the adjacency between two intersections and an approach connects two intersections. 
Among all intersections $\mathcal{V}=\{\mathcal{V}_{internal}\cup \mathcal{V}_{external}\}$, $\mathcal{V}_{internal}$ stands for the intersection whose traffic signals are considered to be controlled and $\mathcal{V}_{external}$ can be seen as sinks or sources of traffic flows. Approaches are further categorized into three types based on the type of the starting and ending intersections. If the starting intersection is external, then the approach is an entry approach in $\mathcal{E}_{entry}$. If the ending intersection is external, then the approach is an exit approach in $\mathcal{E}_{exit}$. If both the starting and ending intersections are internal, then the approach is an internal approach in $\mathcal{E}_{internal}$.
The neighborhood of intersection $v$ is denoted as $NB_v=\{u|(v,u)\in \mathcal{E}_{internal}\}$. 

An approach $e_{vu}$ serves as the incoming route where vehicles enter intersection $u$ and as the outgoing route where vehicles exit intersection $v$. Each approach $e$ includes multiple lanes, referred to as $L[e]$. All lanes on the same approach are adjacent lanes.
Then, we have incoming and outgoing lanes corresponding to different approaches. $In_v$ denotes the set of incoming lanes of intersection $v$ and $Out_v$ denotes the set of outgoing lanes of intersection $v$. Then, all incoming lanes of the traffic network is $\mathcal{L}_{in}=\cup_{v \in \mathcal{V}_{internal} } In_v$.

\begin{figure}[t]
    \centering
    \includegraphics[width=0.5\textwidth]{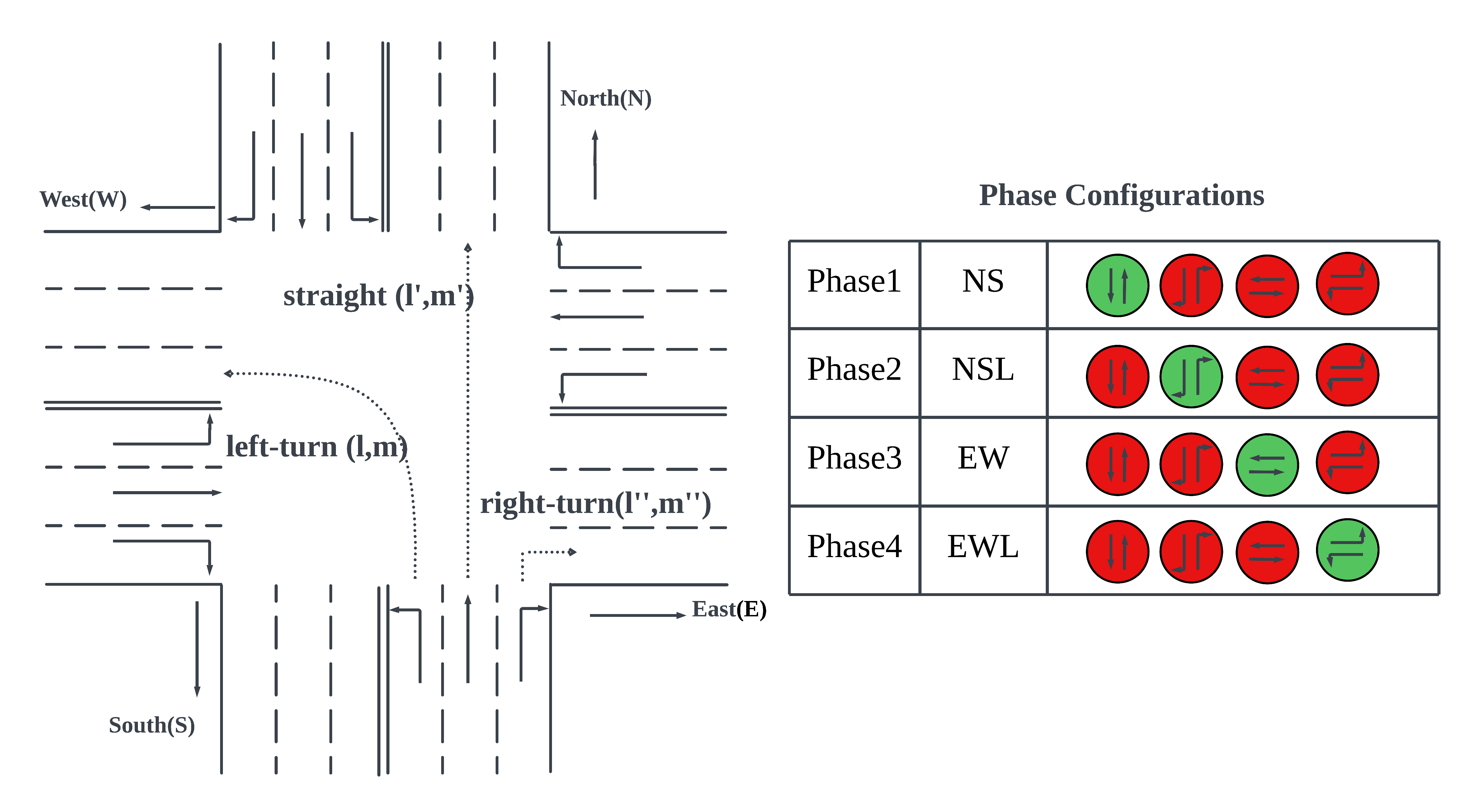}
    \caption{Isolated Intersection and Phase Configuration}
    \label{fig:Traffic network}
\end{figure}

A traffic movement $(l,m)$ at intersection $v$ is defined as a pair of one incoming lane $l \in In_v$ and one outgoing lane $m \in Out_v$. A phase is a set of permitted or restricted traffic movements. As illustrated right side in Fig. \ref{fig:Traffic network}, one intersection has four phases which are North-South Straight (NS), North-South Left-turn (NSL), East-West Straight (EW), and East-West Left-turn (EWL); right-turn movements are always permitted.

\begin{table}[]
    \centering
    \caption{Notion Table}
    \begin{tabular}{|c|c|}
    \hline
        $\mathcal{V}$ & set of all intersections \\
        \hline
        $\mathcal{E}$ & set of all approaches\\
        \hline
        $NB_{v}$ & neighborhood intersections of $v$\\
        \hline 
        $In_v$ & set of all incoming lanes of intersection $v$\\
        \hline 
        $Out_v$ & set of all outgoing lanes of intersection $v$\\
        \hline
        $x(l,m)$ &number of vehicles leaving $l$ to $m$\\
        \hline
        $x(l)$ & number of vehicles on lane $l$\\
        \hline 
        $c(l)$ & discharging rate of lane $l$\\
        \hline
        $a(l)$ & 1 if the signal for lane $(l)$ is green;0 otherwise\\
        \hline
        $r(l,l')$ & routing proportion from lane $l$ to its adjacent lane $l'$\\
        \hline
    \end{tabular}

    \label{tab:Notion Table}
\end{table}

\subsection{Markov Grame Framework and Q-Learning}
The multi-agent reinforcement learning problem is typically modeled as a Markov Game (MG) \cite{littman1994markov}, defined as a tuple$\langle \mathcal{N},\mathcal{S},\mathcal{O},\mathcal{A}, R, P,\gamma \rangle$ where $\mathcal{N}$ represents the set of agents, $\mathcal{S}$ denotes the state space, $\mathcal{O}=\{\mathcal{O}_1,...,\mathcal{O}_{|\mathcal{N}|}\}$ denotes the space of local observations for each individual agent $i$ and each local observation is generated partially from $\mathcal{S}$, $\mathcal{A}=\{\mathcal{A}_1,...,\mathcal{A}_{|\mathcal{N}|}\}$ denotes the set of joint action space. The local reward function $R_i \in R: \mathcal{O} \times \mathcal{A}\rightarrow \mathbb{R}$ maps a pair of observation and joint action to a real number. The transition probability $P: \mathcal{S}\times \mathcal{A} \times\mathcal{S}\rightarrow [0,1] $ assigns a probability to each state-joint action-state transition. $\gamma$ denotes the reward discounted factor which manages the trade-off between immediate and future rewards. 

Each agent $i$ in MG has its own policy $\pi_i$ : $\mathcal{O}_i \times \mathcal{A}_i \rightarrow [0,1]$ indicating the probability distribution of its action over the observation of agent $i$. Each agent tries to maximize its own expected cumulative reward, i.e., the state value function
\begin{equation}
    V(o_i)=\mathbb{E}_{\pi_i}[\sum_{k=0}^\infty \gamma^k r_{i,t+k} |o_{i,t}=o_i]
\end{equation} and the Q-value function
\begin{equation}
    Q(o_i,a_i)=\mathbb{E}_{\pi_i}[\sum_{k=0}^\infty \gamma^k r_{i,t+k} |o_{i,t}=o_i,a_{i,t}=a_i)]
\end{equation}

Traditional tabular Q-learning method stores Q-value in a table\cite{watkins1992q}. However, for some complicated problems with large state space and action space, tabular Q-learning becomes computationally impractical. Deep Q-network (DQN) utilizes a neural network to approximate Q-value and utilizes gradient descent to update the parameters\cite{mnih2015human}.
The loss function for DQN is 
\begin{equation}
    L(\theta_i)=\mathbb{E}_{(o_{i,t},a_{i,t},r_{i_t},o_{i,t+1})\sim D}[(y_{i,t}-Q(o_{i,t},a_{i,t};\theta_i))^2]
\end{equation}
where 
\begin{equation}
    y_{i,t}=r_{i,t}+\gamma \max_{a'}Q(o_{i,t+1},a';\theta_i^-)
\end{equation}
$\theta_i$ denotes the parameter of DQN, $\theta_i^-$ denotes the parameter of the target DQN and $D$ is the experience buffer.
\subsection{GAT}
\label{sec: GAT}
The GAT was proposed to capture hidden features for data in the forms of graphs\cite{velivckovic2017graph}. 
The input of the single-head GAT is a set of features with nodal structure, $\mathbf{h}=\{h_1,h_2,...,h_N\}$, $h_i\in \mathbb{R}^F$, where $N$ is the number of nodes and $F$ is the number of features in each node. The output of the layer is a set of node features, $\mathbf{h}'=\{h_1',h_2',...,h_N'\}$, $h_i'\in \mathbf{R}^{F'}$. The first step is to compute the correlated importance coefficients $ \mathbf{E}\in \mathbb{R} ^{N\times N}$ between nodes by embedding the input features into a higher dimension using a shared weight matrix $\mathbf{W}\in \mathbb{R}^{F'\times F}$ followed by a self-attention mechanism, i.e., 
\begin{equation}
    e_{ij}=\text{LeakyReLU}(a^T[\mathbf{W}h_i||\mathbf{W}h_j])
\end{equation}
where $a\in \mathbb{R}^{2F'}$ and $||$ is the concatenation operation.
Next, a masked attention mechanism is applied to allow each node only to consider the importance coefficients among its neighboring node. The selected importance coefficients are then normalized by the softmax function
\begin{equation}
    \mathbf{\alpha}= \text{exp}(\mathbf{E}) \oslash (\text{exp}(\mathbf{E}) \cdot \mathbf{M})
\end{equation}
where $\oslash$ is the element-wise division operation between matrices and $\mathbf{M}$ is the adjacent matrix of these nodes, i.e.,
\begin{equation}
    \mathbf{M}_{ij}=\begin{cases}
        1 & \text{ $i$ and $j$ are neighbourhoods}\\
        0 & \text{ otherwise}
    \end{cases}
\end{equation}
. Once we have normalized importance coefficients, the final hidden feature of each node is a weighted linear combination of the embedded features of its neighboring nodes, i.e.,
\begin{equation}
    h'_i=\sum_{j \in N_i}\alpha_{ij}\mathbf{W}h_j
\end{equation} where $N_i$ is the neighborhood of node $i$.
However, the single-head GAT could be unstable in certain circumstances. Therefore, the multi-head GAT is introduced to stabilize the learning process. Compared to single-head attention, multi-head attention applies $K$ independent attention mechanism with $K$ independent pairs of $a^k$ and $\mathbf{W}^k$ involved. The input of the multi-head attention layer remains unchanged while the output of that is a concatenation of each single-head attention's results, i.e.,
\begin{equation}
    h'_i=\mathbin\Vert_{k=1}^{K}\sum_{j \in N_i}\alpha^k_{ij}\mathbf{W}^kh_j
\end{equation} 
and $h_i'\in \mathbb{R}^{KF'}$. To conclude, the whole process of the multi-head attention layer is denoted as 
\begin{equation}
    \mathbf{h'}=\text{GAT}(\mathbf{h},M)
\end{equation}
where $M$ is the neighborhood matrix for mask attention

\section{Markovian Property of RL-based Regional Traffic Signal Control Process}
\label{sec: regional markovian property}
A store-and-forward queueing network model is proposed to model the transition of the state of a single intersection and is used to prove such transition satisfies the property of Markov chain\cite{wei2019presslight,varaiya2013max}. We first revisit the single intersection queueing network model by specifying lane-to-lane movements and adjacency lanes. Then, we extend the queueing network model to a group of intersections. 
\begin{figure*}[t]
    \centering
    \includegraphics[width=\textwidth]{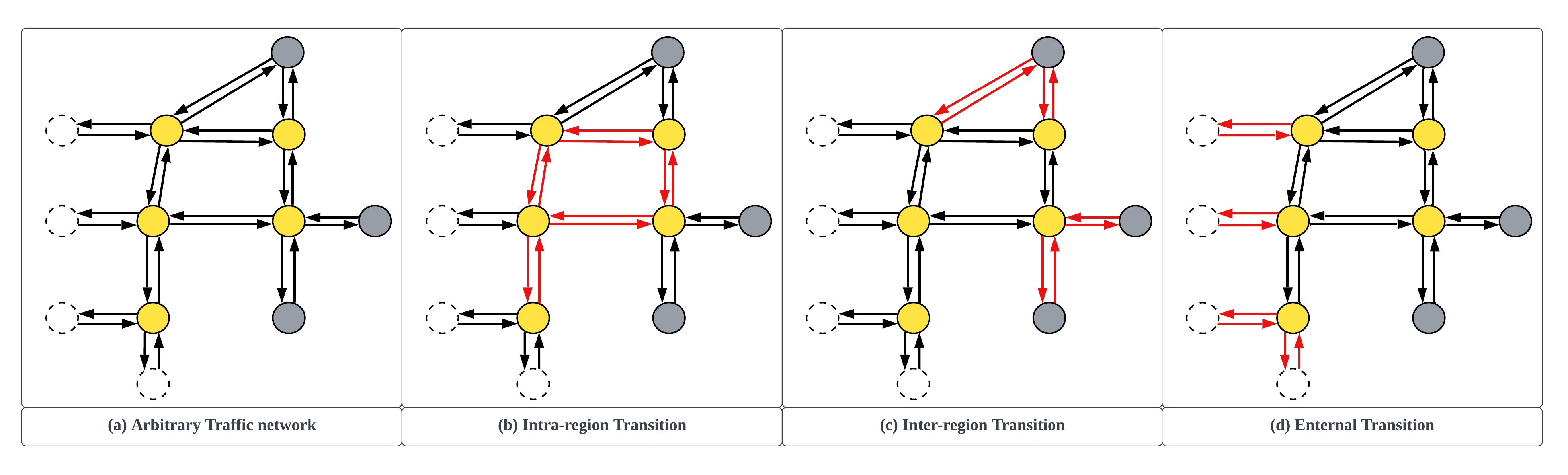}
    \caption{An arbitrary traffic network with external intersections (cycles with dash line) and internal intersections (cycles with solid line). Intersections of the current region $\mathcal{W}$ are filled with \textcolor{YellowOrange}{yellow} and intersections outside the current region are filled with \textcolor{gray}{gray}. Intersections that are not adjacent to $\mathcal{W}$ are omitted for simplicity. \textcolor{red}{Red} arrows in (b), (c), and (d) describe specific interactions in three scenarios. }
    \label{fig: transition_example}
\end{figure*}
\subsection{Single Traffic Signal Control Modeling}

For internal intersection $v\in \mathcal{V}_{internal}$ and its incoming lane $l$, the number of vehicles leaving $l$ to an outgoing lane $m$ at the beginning of period $t$ is denoted as $x^t(l,m)$. For simplicity, we omit $m$ as each lane $l$ has one unique downstream lane. Two variables independent of $x^t(l)$ are defined follows:
\begin{itemize}
    \item Routing proportion $r^t(l,l')$: After a vehicle enters an incoming lane, it can either stay or change to an adjacent lane for the next movement. Therefore, for an incoming lane $l$, a non-negative i.i.d random variable $r(l,l')$ denotes the proportion of the entering vehicles that move to lane $l'$ from lane $l$. The sum of routing proportion from $l$ to all lanes on approach $e$ which $l$ belongs to is 1, i.e., $\sum_{l'} r^t(l,l')=1$.
    \item Discharging rate $c^t(l,m)$: For each movement $(l,m)$, a non-negative i.i.d random variable $c^t(l,m)$ denotes the queue discharging rate and is bounded by saturation flow rate. 
\end{itemize}

The transition of $x(l)$ in period $(t,t+1)$ involves both entering and leaving vehicles. Entered vehicles are contributed directly by vehicles of the movements from the upstream intersections or by vehicles moved from adjacent lanes. Leaving vehicles will move to lane $m$ if movement $(l,m)$ is permitted, i.e., $a(l)=1$. The queue update equation for an internal lane $l$ on one internal approach $e$ is formulated as follows:

\begin{align}
    &x^{t+1}(l)=x^t(l)\\
+   &\sum_{l' \in Lane[e]}min\{c^t(k',l')\cdot a^t(k'),x^t(k',l') \} \cdot r(l',l)\label{eq: second term}\\
-   &min\{c^{t+1}(l,m)\cdot a^t(l), x^t(l)\} \cdot \textbf{1}(wave(m)\leq wave_{max}(m)) \label{eq: third term}
\end{align}
where $wave(m)$ is the current number of vehicles on lane $m$ and $wave_{max}(m)$ is the capacity of lane $m$. 
The second term (Eq. \ref{eq: second term}) represents the movements of vehicles expected to enter lane $l$. For each lane $l'$ including lane $l$ on incoming approach $e$, there are up to $c(k',l')$ vehicles enter if $a^t(k')=1$. Then, the proportion $r(l',l)$ of vehicles will finally move to lane $l$. The third term (Eq. \ref{eq: third term}) represents the movements of vehicles expected to leave lane$l$ where two conditions must be satisfied. The first condition is that the signal allows the vehicle to pass through the intersection which is $a^t(l)=1$ and the second condition is its downstream lane must have the capacity to take the vehicles which is $wave(m)\leq wave_{max}(m)$.

Similarly, the queue update equation for the entry lane whose upstream intersection is outside of the network is formulated as follows

\begin{align}
    &x^{t+1}(l)=x^t(l)+d^{t+1}(l)\\
-   &min\{c^{t+1}(l)\cdot a^t(l), x^t(l)\} \cdot \textbf{1}(wave(m)\leq wave_{max}(m))
\end{align}
where $d^t(l)$ is the demanding flow from intersection $v\in \mathcal{V}_{external}$.

Since the RL agent generates signal action $a$ and the policy of actions is dependent on state $x$, the queue update equation only depends on state $x$, and the process $X(t)$ is a Markov chain.

\subsection{Regional Traffic Signal Control Modeling}



Based on the single intersection evolution model, we now extend the model for a group of intersections and justify that the process of traffic movements under a group of intersections is still a Markov chain. 

Suppose one region is composed of a group of intersections $\mathcal{W}\subset \mathcal{V}_{internal}$, other intersections are either external intersections or ones with pre-defined behaviors. The incoming and outgoing lanes of these intersections are denoted as $\mathcal{F}_{in}= \cup_{v\in \mathcal{W}} In_v $ and $\mathcal{F}_{out}= \cup_{v\in \mathcal{W}} Out_v$ respectively. 
The state of these intersections is stored in a vector $X(\mathcal{F}_{in})\in \mathbb{R}^{|\mathcal{F}_{in}|}$. $C(\mathcal{F}_{in}) \in \mathbb{R}^{|\mathcal{F}_{in}|}$ is a vector of discharging rate of all lanes in $\mathcal{F}_{in}$.
$A(\mathcal{F}_{in}) \in \mathbb{R}^{|\mathcal{F}_{in}|} $ denoted the signal control phase of all incoming lanes inside the region. $a_{l}= 1$ if the signal of lane $l$ is green and $a_{l}= 0$ otherwise.

\begin{definition}[Movement Matrix]
\label{def movement matrix}
    Movement matrix $\mathbf{M}(\mathcal{F}_1,\mathcal{F}_2)\in \mathbb{R}^{|\mathcal{F}_1|\times|\mathcal{F}_2|}$ between two sets of lanes describes the movements between two sets of lanes $\mathcal{F}_1$ and $\mathcal{F}_2$ where 
\begin{align}
    m(k,l)
    =
    \begin{cases}
        1 & \text{$(k,l)$ is a valid movement}\\
        0 & \text{otherwise}
    \end{cases}
\end{align}

\end{definition}

\begin{definition}[Routing Proportion Matrix]
    Routing proportion matrix $\mathbf{RP}(\mathcal{F}_1,\mathcal{F}_2)\in \mathbb{R}^{|\mathcal{F}_1|\times|\mathcal{F}_2|}$ between two sets of lanes describes the route proportion between two sets of lanes $\mathcal{F}_1$ and $\mathcal{F}_2$ where 
\begin{align}
    rp(l,l')=
    \begin{cases}
        [0,1] & \text{$l, l'$ are adjacent}\\
        0 & \text{otherwise}
    \end{cases}
\end{align}
and 
\begin{align}
    \sum_{l' \in \mathcal{F}_2} rp(l,l')=1 
\end{align}
\end{definition}
\begin{definition}[Blockage Matrix]
Blockage matrix $\mathbf{BM}(\mathcal{F}_1,\mathcal{F}_2)$ between two sets of lanes describes whether the number of vehicles on the downstream lane reaches the lane's capacity where
\begin{align}
    bm(k,l)=
    \begin{cases}
        1 & \text{$(l,k)$ is a movement} \\
          &  \text{and $wave(k)\leq wave_{max}(k)$}\\
        0 & \text{otherwise}
    \end{cases}
\end{align}
. If the capacity is reached, then there is a blockage on the downstream lane and no vehicle can leave from the upstream lanes. 
\end{definition}

The queue updating equation for a region involves movements between intra-region intersections, movements between inter-region intersections, and movements from external intersections, i.e.,
\begin{align}
    \label{eq: region markov eq}
    X^{t+1}(\mathcal{F}_{in})&=X^t(\mathcal{F}_{in})+Intra^t+Inter^t+External^t
\end{align}

\subsubsection{Intra-region}
$Intra^t$ represents the traffic movement caused by the traffic signals inside the region at time $t$ (Fig. \ref{fig: transition_example}(b)). 
\begin{align}
    Intra^t=&\hat{X}^t(\mathcal{F}_{in})\cdot\hat{\mathbf{M}}(\mathcal{F}_{in},\mathcal{F}_{in})\\
    -&\hat{X}^t(\mathcal{F}_{in})\cdot \mathbf{BM}(\mathcal{F}_{in},\mathcal{F}_{out})\\
    =&\hat{X}^t(\mathcal{F}_{in})\cdot(\hat{\mathbf{M}}(\mathcal{F}_{in},\mathcal{F}_{in})-\mathbf{BM}(\mathcal{F}_{in},\mathcal{F}_{out}))
\end{align}
where
\begin{equation}
    \hat{X}^t(\mathcal{F}_{in})=\mathbf{min} \{C(\mathcal{F}_{in})\circ A^t(\mathcal{F}_{in}),X^t(\mathcal{F}_{in})\}
\end{equation} describes the number of vehicles that are about to leave each lane due to the signal, $\mathbf{min}$ denotes the operation of taking element-wise minimum between two vectors, $\circ$ denotes the operation of element-wise multiplication between two vectors,
and
\begin{equation}
\label{eq: rp allocation}
    \hat{\mathbf{M}}(\mathcal{F}_1,\mathcal{F}_{2})=\mathbf{M}(\mathcal{F}_{1},\mathcal{F}_{2}) \cdot \mathbf{RP}^t(\mathcal{F}_{2},\mathcal{F}_{2})
\end{equation}
assigns the vehicles from incoming lanes inside the region to themselves.

\subsubsection{Inter-region}
$Inter^t$ describes the traffic movements caused by the traffic signal outside the region at time $t$   (Fig. \ref{fig: transition_example}(c)). We use $\mathcal{W}'=\cup_{v \in \mathcal{W}} NB_v -\mathcal{W}$ to denote the neighbouring intersections outside the region. 
Then, the number of vehicles coming from outside of the region is denoted as 
\begin{align}
    Inter^t= \hat{X}^t(\mathcal{F}'_{in})\cdot(\hat{\mathbf{M}}(\mathcal{F}'_{in},\mathcal{F}_{in})-\mathbf{BM}(\mathcal{F}'_{in},\mathcal{F}_{in}))
\end{align}.

\subsubsection{External Entry Lane}
Among all incoming lanes in $\mathcal{F}_{in}$, some lanes might belong to $\mathcal{E}_{external}$ which originate from sources (Fig. \ref{fig: transition_example}.(d)). Therefore, the last part $External^t\in \mathbb{R}^{|\mathcal{F}_{in}|}$ represents the vehicles from sources where
\begin{align}
    External^t[l]=
    \begin{cases}
        d_l & \text{ if $l$ originates from sources  }\\
        0 & \text{otherwise}
    \end{cases}
\end{align}
.
Then, the queue updating equation of $X^{t+1}(\mathcal{F}_{in})$ only depends on previous state $X^{t}(\mathcal{F}_{in})$. Therefore, the process of regional traffic signal control is also a Markov chain.

\section{GAT-based 
Communication Technique on lane-level and intersection-level traffic states}
\label{sec: communication strategy}
\begin{figure*}[t]
    \centering
    \includegraphics[width=\textwidth]{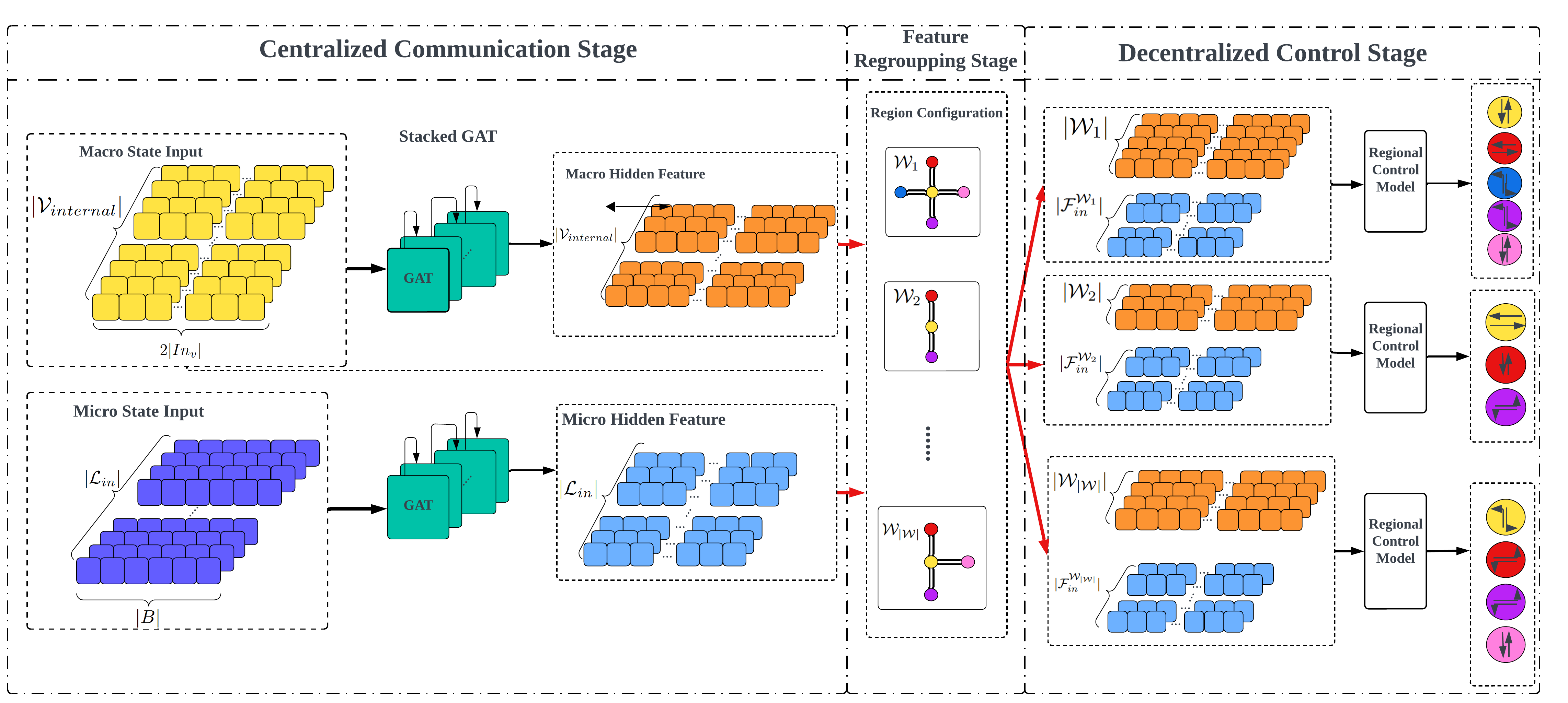}
    \caption{The architecture of proposed information sharing module. Macro and micro state inputs of the whole traffic network are first centralizedly embedded by stacked GATs and two stacks of GATs do not share weights. Then, the macro and micro hidden features are regrouped and concatenated as the observation of each regional agent according to the region configurations respectively. Next, each decentralized agent predicts the best action based on its observations. Finally, the union of the best actions of all agents is the optimal global control strategy for the whole traffic network.}
    \label{fig: GALight}
\end{figure*}
In the queue updating equation of the RTSC process, the transition involves both intra-region and inter-region traffic flows. However, the transition will become non-stationary if we apply multiple RTSC agents in a large traffic network. Then, the control problem turns out to be a Partially observable Markov decision process (POMDP) since the intersections outside the region are controlled by other agents. To alleviate the issue caused by a partially observable environment, we propose a centralized communication module that applies GAT to capture correlations in both macro traffic states among intersections and micro traffic states among lanes. 
\begin{figure}
    \centering
    \includegraphics[width=\linewidth]{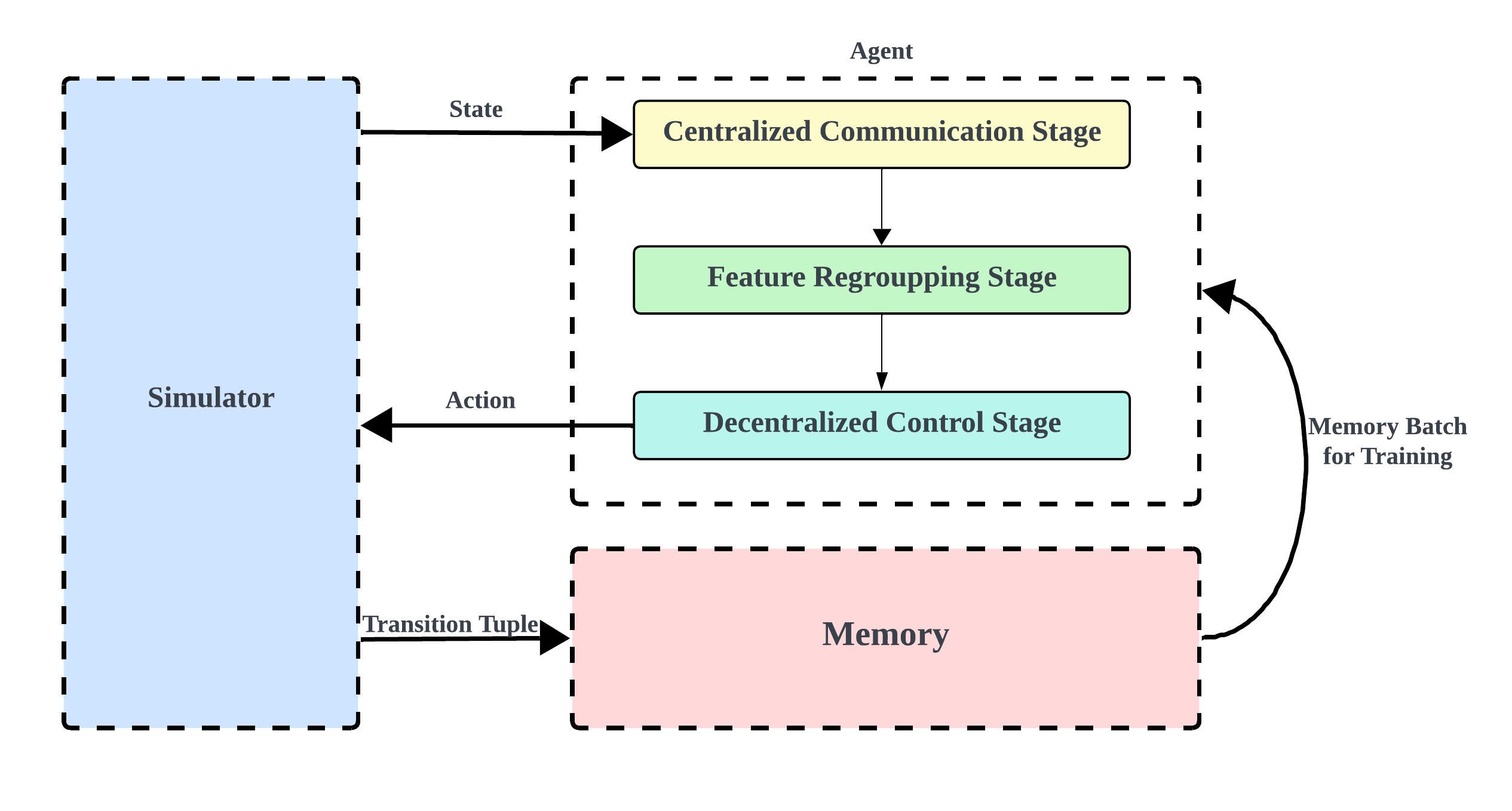}
    \caption{Overall Framework. Our framework contains three components, simulator, agent and memory. The simulator will simulate the traffic environment and offer traffic states for agents. Then agents will make decisions based on traffic states through three stages. First, the centralized communication stage will process network-level traffic state to enable RTSC agents to share information. Then, hidden features will be regrouped and flattened for each region in the feature regrouping stage. Finally, in the decentralized control stage, RTSC agents will choose the actions for their regions in a decentralized manner. The memory component will store the recent transition tuples for future training.}
    \label{fig:framework}
\end{figure}
The overall framework of our work is illustrated in Fig \ref{fig:framework} and the architecture of the proposed communication module is illustrated in Fig. \ref{fig: GALight}. The traffic state of the entire traffic network is first embedded in a centralized manner. Then, the embedded state is split and regrouped as the observation for each RTSC agent and each RTSC agent makes decisions based on its individual observations.

In this section, we first present the formulation of our communication module with two variants and then describe how this module can be aggregated with RegionLight\cite{gu2024large} and RDRL\cite{tan2019cooperative}.

\subsection{Lane-Level GAT}
\label{sec: lanelevel GAT}
\begin{figure}
    \centering
    \includegraphics[width=0.5\textwidth]{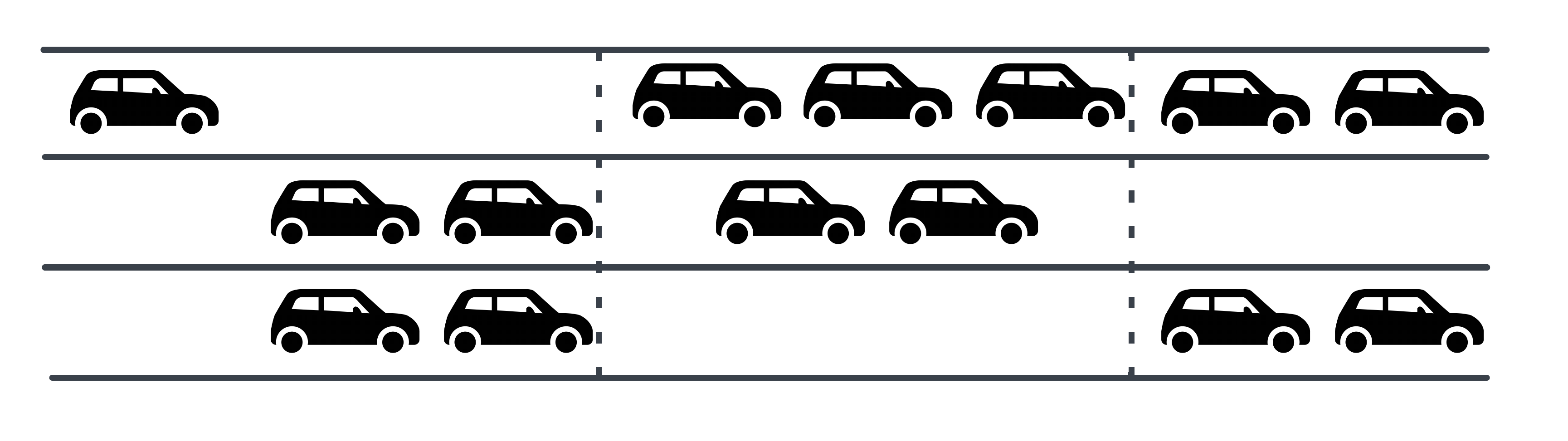}
    \caption{Example of Lane Segmentation. Suppose we have an incoming road with three lanes and each lane is segmented into three cells. Then the lane level state of these incoming lanes is $\{[1,3,2],[2,2,0],[2,0,2]\}$.}
    \label{fig:cell example}
\end{figure}

We segment each incoming lane into $B$ cells with and the number of vehicles traveling inside each cell is observed (Fig. \ref{fig:cell example}). 
Then the input to the lane-level state embedding module is the set of all segmented incoming lanes and is denoted as $\mathbf{S}^{lane}=\{S^{lane}[l],...,S^{lane}[m]\}\in \mathbb{R}^{|\mathcal{L}_{in}|\times B}$, where $S^{lane}[l] \in \mathbb{R}^{B}$ and $l,m \in \mathcal{L}_{in}$. 
Then, inspired by \cite{lin2023temporal}, we propose two movement matrices to describe the neighborhood relationship between lanes and stack several GATs to embed the lane-level states.
\paragraph{Naive Movement Matrix} Based on the Def. \ref{def movement matrix}, vehicles move from one incoming lane to one outgoing lane. Therefore, a correlation exists between the pair of lanes in any valid movement. We propose a naive movement matrix $M^{lane}_{naive} \in \mathbb{R}^{|\mathcal{L}_{in}|\times |\mathcal{L}_{in}|}$ to capture the correlation of both upstream and downstream flow, i.e.,
\begin{equation}
    M^{lane}_{naive}(\mathcal{L}_{in},\mathcal{L}_{in})=\mathbf{M}(\mathcal{L}_{in},\mathcal{L}_{in})+\mathbf{M}^T(\mathcal{L}_{in},\mathcal{L}_{in})+ \mathbf{I}_{|\mathcal{L}_{in}|}
\end{equation}
 where $\mathbf{I}_{|\mathcal{L}_{in}|}$ is the identical matrix of dimension $|\mathcal{L}_{in}|$.
 
\paragraph{Augmented Movement Matrix} The naive movement matrix considers only the traffic movements caused by vehicles passing the intersection. However, in Eq. (\ref{eq: rp allocation}), a vehicle might move to one adjacent lane after it enters one outgoing lane. Therefore, the naive movement matrix fails to consider the lane-changing behaviors of vehicles. Hence, we propose an augmented movement matrix $M^{lane}_{aug}$ to capture the correlation between adjacent lanes more comprehensively, i.e., 
\begin{equation}
    M^{lane}_{aug}(\mathcal{L}_{in},\mathcal{L}_{in})=M^{lane}_{naive}(\mathcal{L}_{in},\mathcal{L}_{in})+ADJ(\mathcal{L}_{in},\mathcal{L}_{in})
\end{equation}
where $ADJ(\mathcal{L}_{in},\mathcal{L}_{in})$ is the adjacent matrix between lanes, i.e.,
\begin{equation}
    adj(l,l')=\begin{cases}
        1 & \text{$l$ and $l'$ are adjacent lanes}\\
        0 & \text{otherwise}
    \end{cases}
\end{equation}. With the help of $ADJ(\mathcal{L}_{in},\mathcal{L}_{in})$, the GATs also compute the importance coefficients between adjacent lanes and consider the evolution between adjacent lanes caused by lane-changing behaviors.

Next, the architecture for lane-level state embedding is listed as follows 
\begin{align}
        \mathbf{h}_1^{lane}&=\text{GAT}(\mathbf{S}^{lane},M^{lane})\\
        \mathbf{h}_2^{lane}&=\text{GAT}(\mathbf{h}_1^{lane},M^{lane})\\
        ...\\
        \mathbf{h}_m^{lane}&=\text{GAT}(\mathbf{h}_{m-1}^{lane},M^{lane})\label{eq: last hidden micro}
\end{align}
where $M^{lane}$ denotes the mask attention matrix between lanes.

\subsection{Intersection-Level GAT}
The input to the intersection-level state embedding module is a summarised information of each internal intersection $v \in \mathcal{V}_{internal}$ and is denoted as  $\mathbf{S}^{itsx}=\{S^{itsx}[v],...,S^{itsx}[u]\}\in \mathbb{R}^{|\mathcal{V}_{internal}|\times 2|In_v|}$ where
\begin{equation}
    S^{itsx}[v]=\{\{wait(l)\}_{l \in In_v},\{wave(l)\}_{l \in In_v} \}
\end{equation} and $wait(l)$ is the number of waiting vehicles on lane $l$. Note that, if the number of cells in the lane-level state is set to 1, then the lane-level state is equivalent to the wave on each lane. 

Similar to previous work\cite{wei2019colight}, we use the adjacent matrix between intersections as the masked attention matrix in GAT, i.e., 
\begin{align}
        \mathbf{h}_1^{itsx}&=\text{GAT}(\mathbf{S}^{itsx},M^{itsx})\\
        \mathbf{h}_2^{itsx}&=\text{GAT}(\mathbf{h}_1^{itsx},M^{itsx})\\
        ...\\
        \mathbf{h}_m^{itsx}&=\text{GAT}(\mathbf{h}_{m-1}^{itsx},M^{itsx}) \label{eq: last hidden macro}
\end{align}
where $M^{itsx}$ is the adjacent matrix between intersections and $ m^{itsx}(v,u)=1$ if $(v,u)\in \mathcal{E}_{internal}$; $ m^{itsx}(v,u)=0$ otherwise. 

\subsection{Intersection Grouping and Observation Construction}
The hidden features of lane-level and intersection-level traffic states are then regrouped according to the configuration of each region. Each region $\mathcal{W}_i$ is a set of intersections that obeys the following two constraints:
\begin{align}
    \cup_i \mathcal{W}_i =\mathcal{V}_{internal}\\
    \mathcal{W}_i \cap \mathcal{W}_j= \emptyset, \forall i\neq j
\end{align}
where the first constraint ensures all regions cover all internal intersections and the second constraint ensures all regions are disjoint. In this paper, we follow the constrained network partition rule proposed in \cite{gu2024large} to construct the configuration of each region but with the dummy intersection removed. Then, each region $\mathcal{W}_i$ contains at least one internal intersection $v$ and a subset of the neighborhoods of $v$, i.e., 
$\mathcal{W}=\{\{v\}\cup U\}$ where $U\subset NB_v$.
\begin{remark}
One GAT embeds each node feature with its neighborhood. With more stacked GAT, each hidden node feature is embedded in more nodes. Since the diameter of each region is at most 2, we set the number of stacked GAT to 2, i.e., $m=2$ in Eq.(\ref{eq: last hidden micro}) and (\ref{eq: last hidden macro}). Hence, the complexity analysis of both modules can be analyzed.
For lane-level GAT, the computational complexity for single-head is $\mathbf{O}(|\mathcal{L}_{in}||\mathbf{h}_1^{lane}[l]||\mathbf{h}_2^{lane}[l]|+|g||\mathbf{h}_2^{lane}[l]|)$ where $g$ denotes the number of one in $ M^{lane}_{naive}(\mathcal{L}_{in},\mathcal{L}_{in})$ or $ M^{lane}_{aug}(\mathcal{L}_{in},\mathcal{L}_{in})$
For intersection-level GAT, the computational complexity for single-head is $\mathbf{O}(|\mathcal{V}_{internal}||S^{itsx}[v]||\mathbf{h}_2^{itsx}[v]|+|\mathcal{E}_{internal}||\mathbf{h}_2^{itsx}[v]|)$.
Although we apply multi-head GAT, multi-head computation is independent and can be parallelized.

\end{remark}
Next, the hidden features $\mathbf{h}_m^{lane}$ and $\mathbf{h}_m^{itsx}$ are regrouped and concatenated as the observation feature for regional control agents, i.e.,
\begin{equation}
    O_i=\{\{\mathbf{h}_m^{lane}[l]\}_{l \in In_v},\mathbf{h}_m^{itsx}[v]\}_{v\in \mathcal{W}_i}
\end{equation}
where $\{\mathbf{h}_m^{lane}[l]\}_{l \in In_v}$ represents the hidden micro traffic state for all incoming lanes to the region and $\{ \mathbf{h}_m^{itsx}[v]\}_{v\in \mathcal{W}_i}$ represents the hidden macro traffic state for all intersections in the region. Note that in different networks, the number of intersections in different regions can be different, indicating that the dimensions of observation for each agent can differ. Therefore, two possible strategies can be applied. One naive strategy is to model each agent specifically according to the exact configuration of each region which indicates that agents do not share parameters of their networks. The other strategy is to follow the modeling in \cite{gu2024large} which is to fix the maximum number of intersections inside one region and the maximum number of lanes one intersection can have, followed by using dummy intersections to fill the absence.

\subsection{Action Space and Reward Function}
As defined in Fig. \ref{fig:Traffic network}, each intersection has four phases: North-South through (NS), East-West through (EW), North-South left-turn, and East-West left-turn (EWL). Therefore, the joint action space for each region is denoted as 
\begin{equation}
    \mathcal{A}_i=\{NS,NSL,EW,EWL\}^{|\mathcal{W}_i|}
\end{equation}.
The ultimate goal for TSC is to reduce the average travel time of all vehicles. However, this delayed metric is not directly applicable to the DRL problem as agents need immediate rewards to optimize performance.  
The reward function of each agent is the negative sum of the number of waiting vehicles on all incoming lanes inside the region, i.e., 
\begin{equation}
    R_i^t=-\sum_{v \in \mathcal{W}_i}\sum_{l \in In_v}wait^t(l)
\end{equation} same as the reward function in previous work\cite{gu2024large,zheng2019diagnosing,zhang2022neighborhood}.
\subsection{Centralized Experience Buffer and $\epsilon$-greedy policy}
In common MADRL training, previous transition tuples are sampled from its experiment randomly and are used by agents to update parameters independently. However, given that the proposed communication modules embed macro and micro traffic states in a centralized manner, this independent sampling technique can lead to stability issues. To address the issues and ensure that our communication modules effectively capture the correlation between traffic states, we store the experiences of all agents in a centralized experience buffer. Sampling experiences from this centralized buffer will maintain the underlying traffic dynamics between intersections while facilitating information exchange.

To achieve a trade-off between exploration and exploitation, agents follow $\epsilon$-greedy policy\cite{watkins1992q}, i.e.,
\begin{equation}
    \pi(s)=\begin{cases}
        \text{Random Action} & \text{with probability $\epsilon$}\\
        argmax_{a\in A} Q(s,a) & \text{with probability $1-\epsilon$}
    \end{cases}
\end{equation}.
The value of $\epsilon$ starts from $\epsilon_{max}$ close to 1 and decays to $\epsilon_{min}$ in certain steps.

\section{Experiment and Result}
\label{sec: experiment and result}
In this section, we examine the performance of our communication modules when applied to RegionLight and RDRL, specifically GA2-RegionLight(Naive), GA2-RegionLight(Aug), GA2-RDRL(Naive), and GA2-RDRL(Aug). These four aggregated models are deployed in both real and synthetic traffic scenarios. We compare their performance against other baseline models and present the improvements observed across all regions. Additionally, to assess the robustness and stability of our communication modules, we test various sets of hyperparameters.

\subsection{Experiment Settings}
\begin{figure}[htp]
    \centering
    \subfloat[Hangzhou]{
    \includegraphics[width=0.3\textwidth]{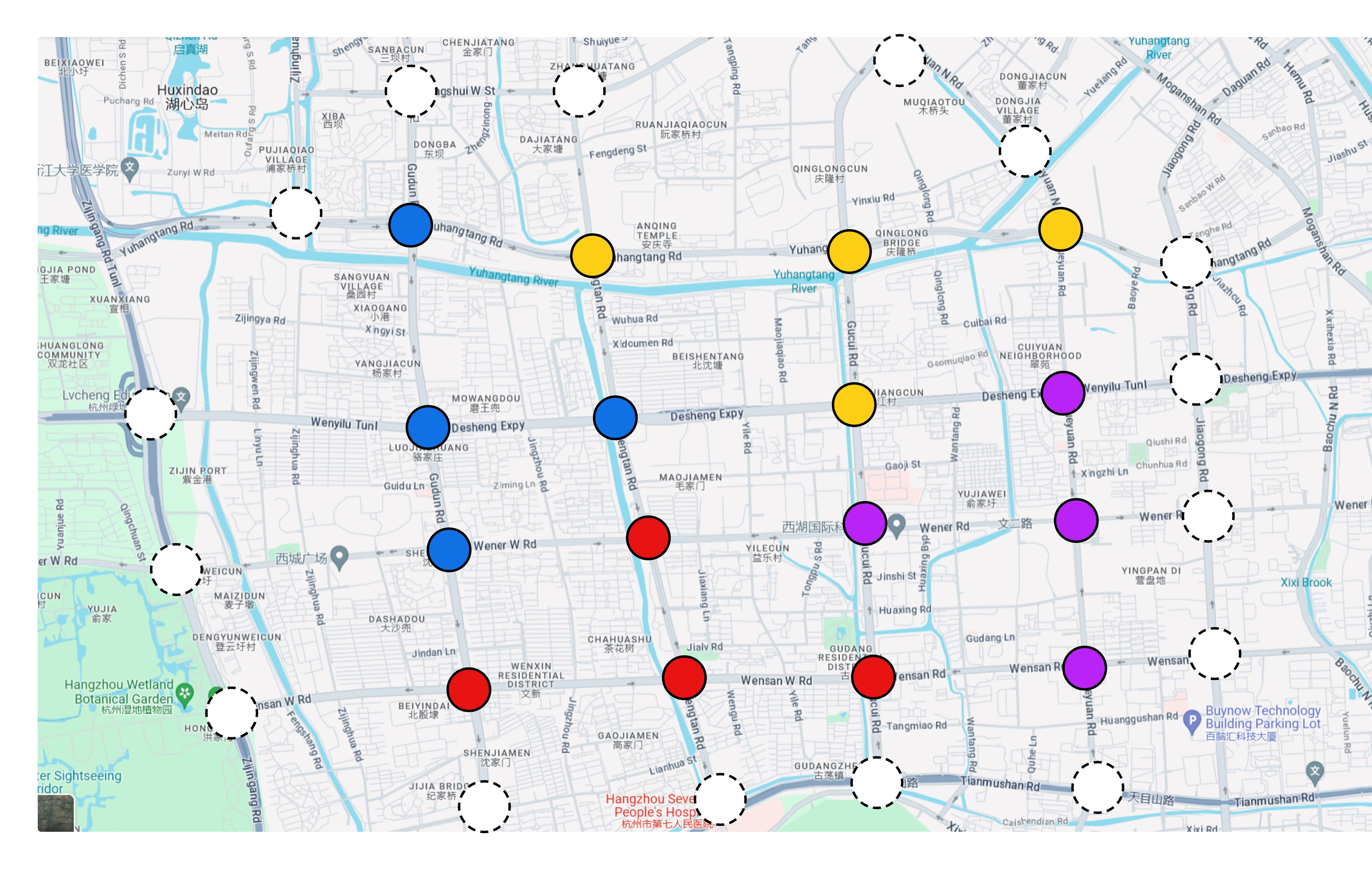}
    }
    \hfill

    \centering
    \subfloat[New York]{
    \includegraphics[width=0.3\textwidth]{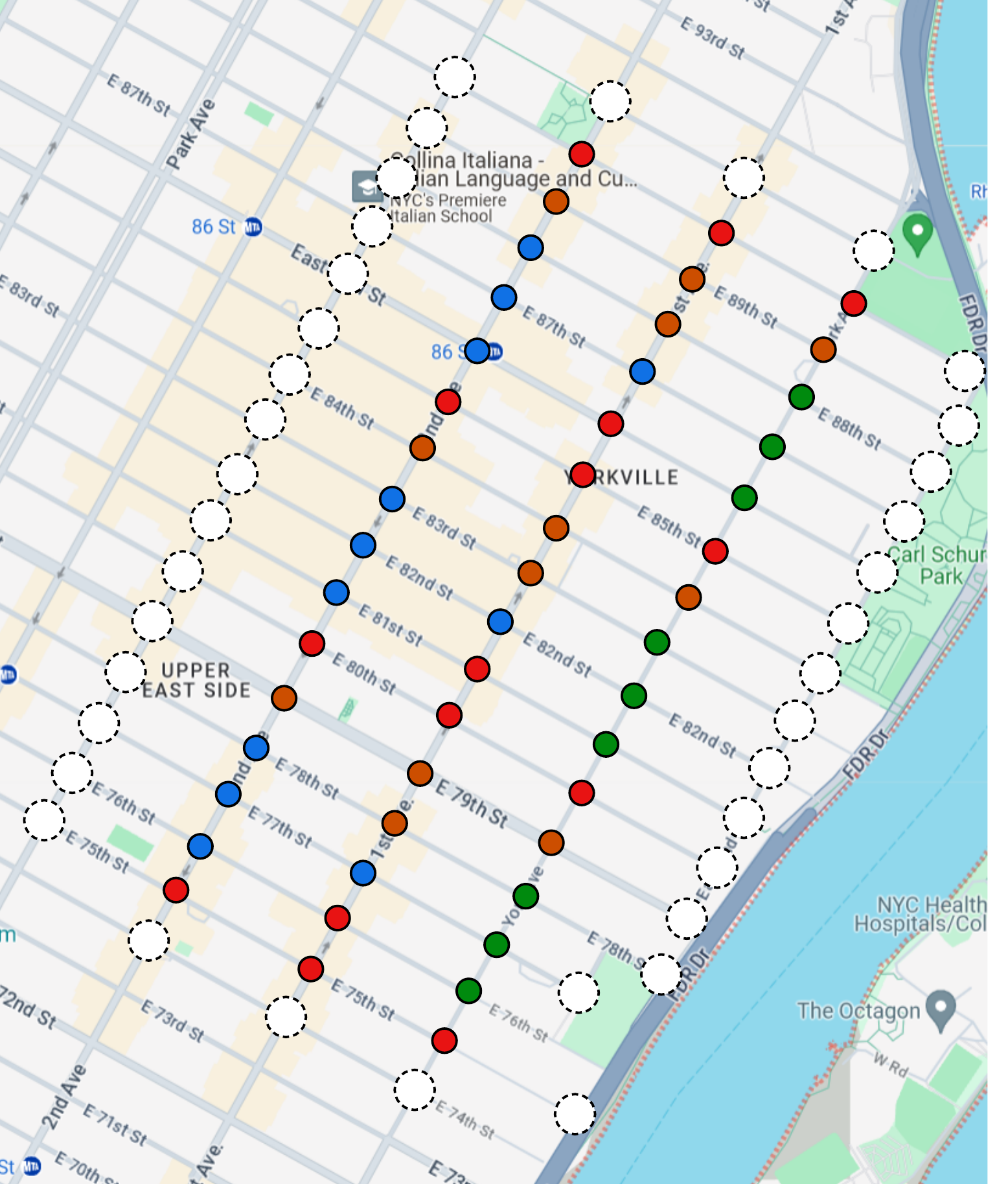}
    }
    \hfill
    \caption{Traffic Grid Network for Hangzhou and New York scenarios. Intersections are annotated by dots. Dots with solid borders indicate internal intersections while dots with dashed borders indicate external intersections. Additionally, we use different colors to distinguish different regions. For example, the $4\time4$ Hangzhou network is partitioned into 4 regions, and the $16\time 3$ New York network is partitioned into 13 regions. }
    \label{fig:hangzhou and manhattan}
\end{figure}

In our experiment, we utilized three grid traffic networks which are Hangzhou ($4\times 4$), synthetic ($4\times 4$), and New York ($16\times 3$). The illustrations of two real traffic networks (Hangzhou and New York) are shown in Fig.\ref{fig:hangzhou and manhattan}. In the Hangzhou grid, we applied both a flat traffic flow and a peak traffic flow. The volumes of these two traffic flows were collected from camera data in Hangzhou city, while the turning ratios were synthesized from taxi GPS data statistics. In the Hangzhou scenarios, the average turning ratios for vehicles are distributed as follows: 10\% turning left, 60\% going straight, and 30\% turning right but the exact turning ratios at different intersections are not identical. Similarly, the traffic flow for the New York scenario is sampled from taxi trajectory data. The traffic flow for the synthetic scenario is sampled from a Gaussian distribution with a mean of 500 vehicles/hour/lane. The datasets are open-source\footnote{https://traffic-signal-control.github.io/\#open-datasets} and some statistical information is listed in Table \ref{tab: traffic info}.

\begin{table}[]
    \centering
    \caption{The Configuration of Scenarios}
    \begin{tabular}{|c|c|c|c|}
         \hline
         Scenario& $|\mathcal{V}_{internal}|$ &Approach Length & Throughput\\
         \hline
         Hangzhou(Flat)& 16& 800m(EW),600m(NS)&2983\\ 
         \hline
         Hangzhou(Peak) & 16&800m(EW),600m(NS)&6538\\
        \hline
        Synthetic & 16& 300m(EW),300m(NS)&11231\\
        \hline
        New York & 48& 350m(EW),100m(NS)&2824\\
        \hline
    \end{tabular}

    \label{tab: traffic info}
\end{table}

An open-source traffic simulator CityFlow\cite{zhang2019cityflow} is selected to simulate the above scenarios. For each scenario, we simulate 4000 time steps, with each time step representing one second in the real world. The agent selects one action every 20 time steps. An all-red phase, lasting 3 time steps, is inserted between two different phases to clear the intersection and ensure safety. The length of each episode is 200.

\subsection{Baseline Model and Hyperparameter}
We choose both conventional TSC methods and RL methods as baselines.

Conventional TSC methods:
\begin{itemize}
    \item Fixed time: Fixed time is a classic TSC method that switches traffic signal phases according to a predetermined schedule and is not influenced by real-time traffic conditions.
    \item Self-organizing traffic lights (SOTL)\cite{gershenson2004self}: SOTL considers real-time traffic demands of all phases. The switch will be approved if the number of approaching vehicles in the current phase reaches a threshold.
\end{itemize}

Offline RL methods first collect the experience of each entire episode and then perform learning after the episode terminates:
\begin{itemize}
    \item CoLight\footnote{https://github.com/wingsweihua/colight}: The state of CoLight consists of a one-hot vector of the current phase and the number of vehicles on each incoming lane. The reward for each intersection is the sum of all waiting vehicles on all incoming lanes.
    \item Efficient-PressLight\footnote{https://github.com/LiangZhang1996/Efficient\_XLight}\cite{wu2021efficient}: Efficient Pressure extends the original computation of pressure presented in \cite{varaiya2013max} by considering lane-changing behaviors on outgoing lanes. This modified pressure calculation is then used as the reward function.
    \item Efficient-CoLight\cite{wu2021efficient}: Similarly to Efficient-PressLight, this method uses efficient pressure as the reward function.
\end{itemize}

Online RL methods learn from existing during the interaction with the environment:
\begin{itemize}
    \item ENC-HDQN\footnote{https://github.com/RL-DLMU/PNC-HDQN}: ENC-HDQN assumes that if two intersections are positively correlated to the number of vehicles between two intersections. Then, an empirical threshold is defined to 
    divide the correlation into three categories.
    \item PNC-HDQN: PNC-HDQN stores synchronous reward trajectories for all agents. Then, the correlation degree is computed using Pearson coefficient of these trajectories.
    
    \item R-DRL: R-DRL applies deep deterministic policy gradient\cite{lillicrap2015continuous} to predict proto-action and then use Wolpertinger Architecture\cite{dulac2015deep} to map continuous actions to candidate discrete actions.
    
    \item RegionLight\footnote{https://github.com/HankangGu/RegionLight}: Extended from the Branching Dueling Q-network (BDQ) \cite{tavakoli2018action}, Adaptive-BDQ (ADBQ) is proposed to mitigate the negative effects of fictitious intersections introduced during network partitioning. Unlike R-DRL, ADBQ directly predicts discrete actions for each region.
\end{itemize}
To ensure fairness, we run the source code of baselines and the code of the proposed method is open-sourced\footnote{https://github.com/HankangGu/GA2NaiveAug}. The hyperparameters of the RegionLight and R-DRL follow 
the original paper\cite{gu2024large,tan2018large}, and the hyperparameters for our GA2 module and experiment are listed in Table \ref{tab: hyperparameter}.
\begin{table}[htp]
    \centering
        \caption{Hyperparameter Configuration}
    \begin{tabular}{|c|c|c|}
    \hline
         Module& Name & Value  \\
         
        \hline
        \multirow{5}*{GA2}& $L_2$ regularization  & 0.0005\\\cline{2-3}&Activation Function&Leaky-ReLu\\
        \cline{2-3}&Head Num for GAT &8\\    
        \cline{2-3}&Hidden Unit for $1_{st}$ GAT &8\\  
        \cline{2-3}&Hidden Unit for $2_{nd}$ GAT &16\\ 
         \hline
         \multirow{8}*{RL}& $\gamma$ &0.9  \\ 
         \cline{2-3}&Cell Number Size&5\\
         \cline{2-3}&Replay Buffer Size&200000\\
         \cline{2-3}&Episode Number& 2000\\
         \cline{2-3}&Episode Step Length& 200\\
         \cline{2-3}&Learning Frequency& 5\\
         \cline{2-3}&Target Network Update Frequency &200\\
         \cline{2-3}&Batch Size &32 \\
        \hline
         
         \multirow{3}*{$\epsilon$-greedy Policy}& $\epsilon_{max}$ &1 \\ 
         \cline{2-3}&$\epsilon_{min}$ &0.001\\
         \cline{2-3}& decay steps & 20000\\
                          
         \hline
    \end{tabular}

    \label{tab: hyperparameter}
\end{table}

\subsection{General Results}
Similar to previous works, the performance of TSC techniques is evaluated based on the average travel time of all vehicles. The numerical results of the average travel time for all models are listed in Table \ref{tab: general results} and the best results are marked with red color. From Table \ref{tab: general results}, all MADRL-based models converge in $4\times 4$ networks, while some models fail to converge in the $16\times 3$ network, with the corresponding results omitted in Table \ref{tab: general results}. In general, the average travel time of MADRL-based models is better than Rule-based methods. Among MADRL-based models, the standard deviation of average travel time across different episodes is higher in offline models because learning occurs after the entire simulation episode is completed. Unlike offline models, online models learn during the simulation, allowing them to adjust their policies in real-time based on previous experiences. Among all baselines, although RegionLight deploys the independent RL agents, it achieves the best average travel time in most scenarios except for HangzhouFlat. 

After the two regional signal control models are aggregated with the proposed communication module, the average travel time decreases in all scenarios. Compared to RegionLight among all scenarios, GA2-RegionLight(Naive) improves the average travel time by approximately 5 seconds, while GA2-RegionLight(Aug) improves the average travel time by about 7 seconds. In the Synthetic scenario, the performance of GA2-RegionLight(Naive) is worse than that of RegionLight but the augmented movement matrix fixed this issue. The number of traveling vehicles is significantly higher in the Synthetic scenario compared to other scenarios. Consequently, more frequent lane-changing behaviors occur in this scenario, and the naive movement matrix probably fails to capture the correlation between these behaviors since it ignores the lane-changing behaviors of vehicles.
Compared to RDRL among all scenarios, the average travel time decreased by about 6\% with GA2-Naive and decreased by about 8\% with GA2-Aug.
\begin{table*}[htb]
\renewcommand{\arraystretch}{2}
\centering
\caption{Average Travel Time in Testing Scenarios}
\begin{tabular}{llllll}
\hline
                              & Model                  & Hangzhou(Flat)  & Hangzhou(Peak)  & Synthetic          & New York     \\ \hline
\multirow{2}{*}{Rule-based}   & Fixed  Time            & 799.48        & 1102.67       & 983.83        & 1985.99       \\
                              & SOLT                   & 625.32        & 682.45        & 529.48        & 850.32       \\ \hline
\multirow{3}{*}{Offline}      & CoLight                & 357.63 $\pm$ 4.95  & 476.33 $\pm$ 5.56  & 253.26 $\pm$ 6.23  & 247.92 $\pm$ 5.56  \\
                              & EfficientCoLight       & 349.06 $\pm$ 0.93  & 466.07 $\pm$ 1.67  & 261.49 $\pm$ 4.87  & 279.67 $\pm$ 10.55 \\
                              & EfficientPressLight   & 374.06 $\pm$ 10.20  & 553.59 $\pm$ 15.06 & 289.61 $\pm$ 9.29  & -             \\ \hline
\multirow{4}{*}{Online}       & ENCHDQN                & 428.03 $\pm$ 0.53 & 464.58 $\pm$ 1.54  & 251.36 $\pm$ 1.96  & 242.56 $\pm$ 1.74  \\
                              & PNCHDQN                & 432.57 $\pm$ 1.24  & 459.01 $\pm$ 3.15  & 252.35 $\pm$ 1.88  & -             \\
                              & RDRL                   & 373.33 $\pm$ 3.95  & 490.09 $\pm$ 2.92  & 306.51 $\pm$ 15.40 & 286.03 $\pm$ 5.67  \\ 
                              & RegionLight            & 354.08 $\pm$ 0.97  & 454.37 $\pm$ 2.09  & 249.39 $\pm$ 1.75  & 238.21 $\pm$ 1.43  \\ \hline
\multirow{4}{*}{Online(OURS)} 
                              & GA2-RDRL(Naive)         & \textcolor{red}{344.76} $\pm$ 0.80  & 458.82 $\pm$ 2.82  & 277.14 $\pm$ 1.70  & 254.05 $\pm$ 3.22  \\
                              & GA2-RDRL(Aug)           & 345.22 $\pm$ 0.79  & 453.86 $\pm$ 2.75  &  276.30 $\pm$ 1.15  & 253.83 $\pm$ 3.80   \\ 
                              & GA2-RegionLight(Naive) & 347.75 $\pm$ 0.87  & 451.54 $\pm$ 2.39  & 252.40 $\pm$ 1.60   & 234.40 $\pm$ \textcolor{red}{1.27}   \\
    & GA2-RegionLight(Aug)   & 346.52 $\pm$ \textcolor{red}{0.73}  & \textcolor{red}{449.93} $\pm$ \textcolor{red}{2.14}  & \textcolor{red}{246.84} $\pm$ \textcolor{red}{1.48}  & \textcolor{red}{232.89} $\pm$ 1.29  \\
    \hline
                              
\end{tabular}
\label{tab: general results}
\end{table*}

\subsection{Regional Reward Improvement}
\begin{figure*}
    \centering
    \subfloat[Hangzhou(Flat)]{
    \includegraphics[width=0.24\textwidth]{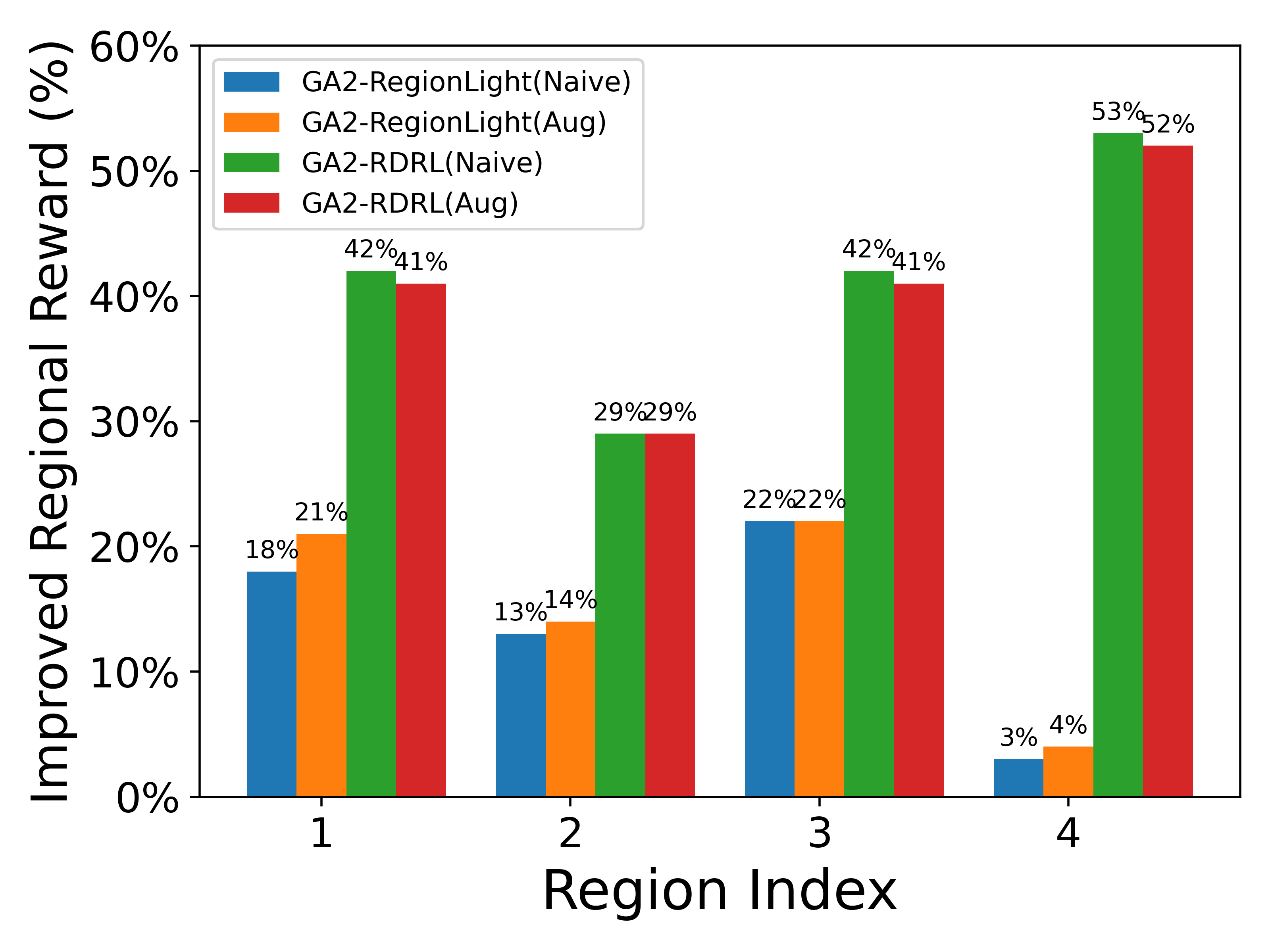}
    }
    \subfloat[Hangzhou(Peak)]{
    \includegraphics[width=0.24\textwidth]{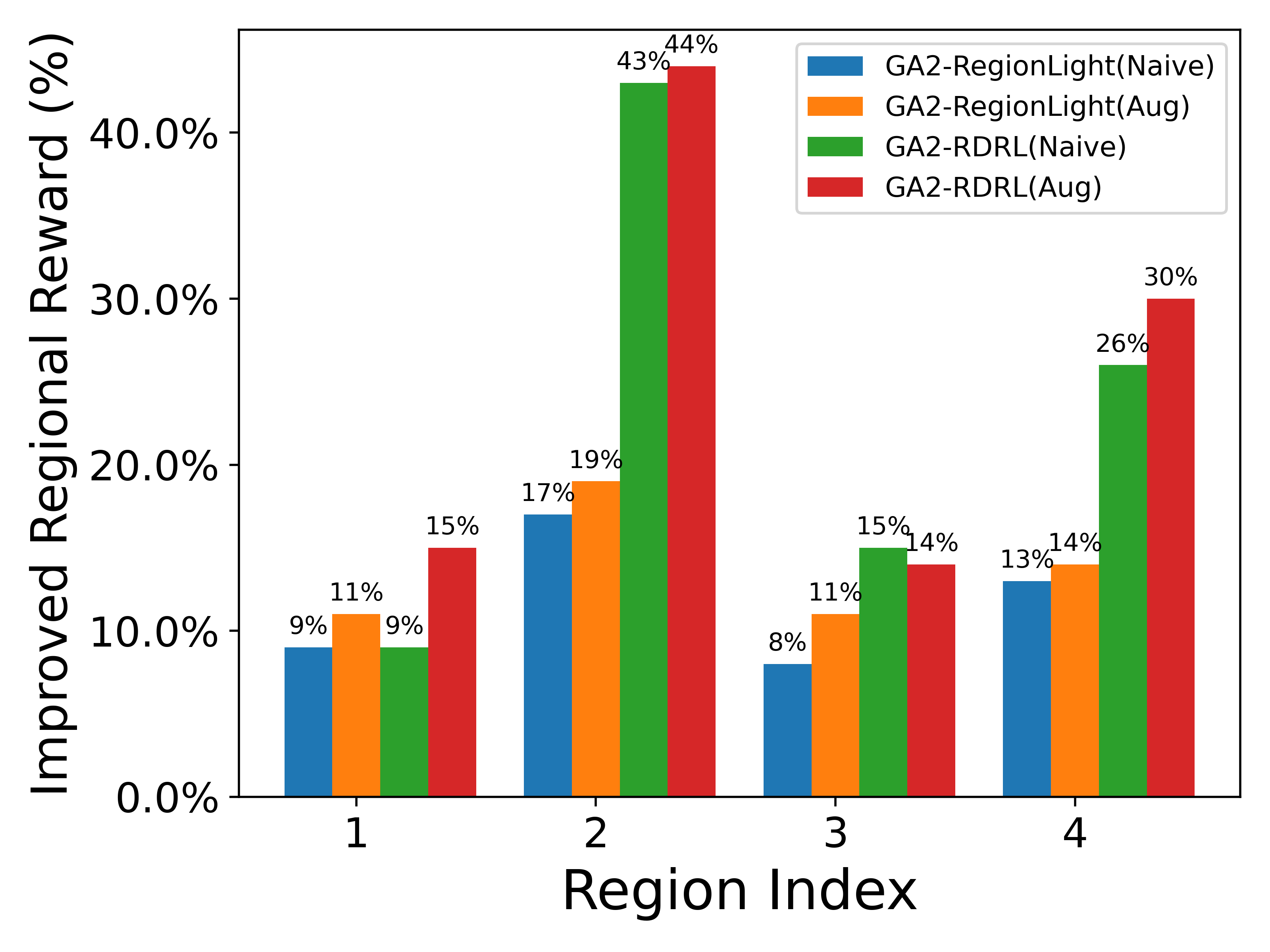}
    }
    \subfloat[Synthetic]{
    \includegraphics[width=0.24\textwidth]{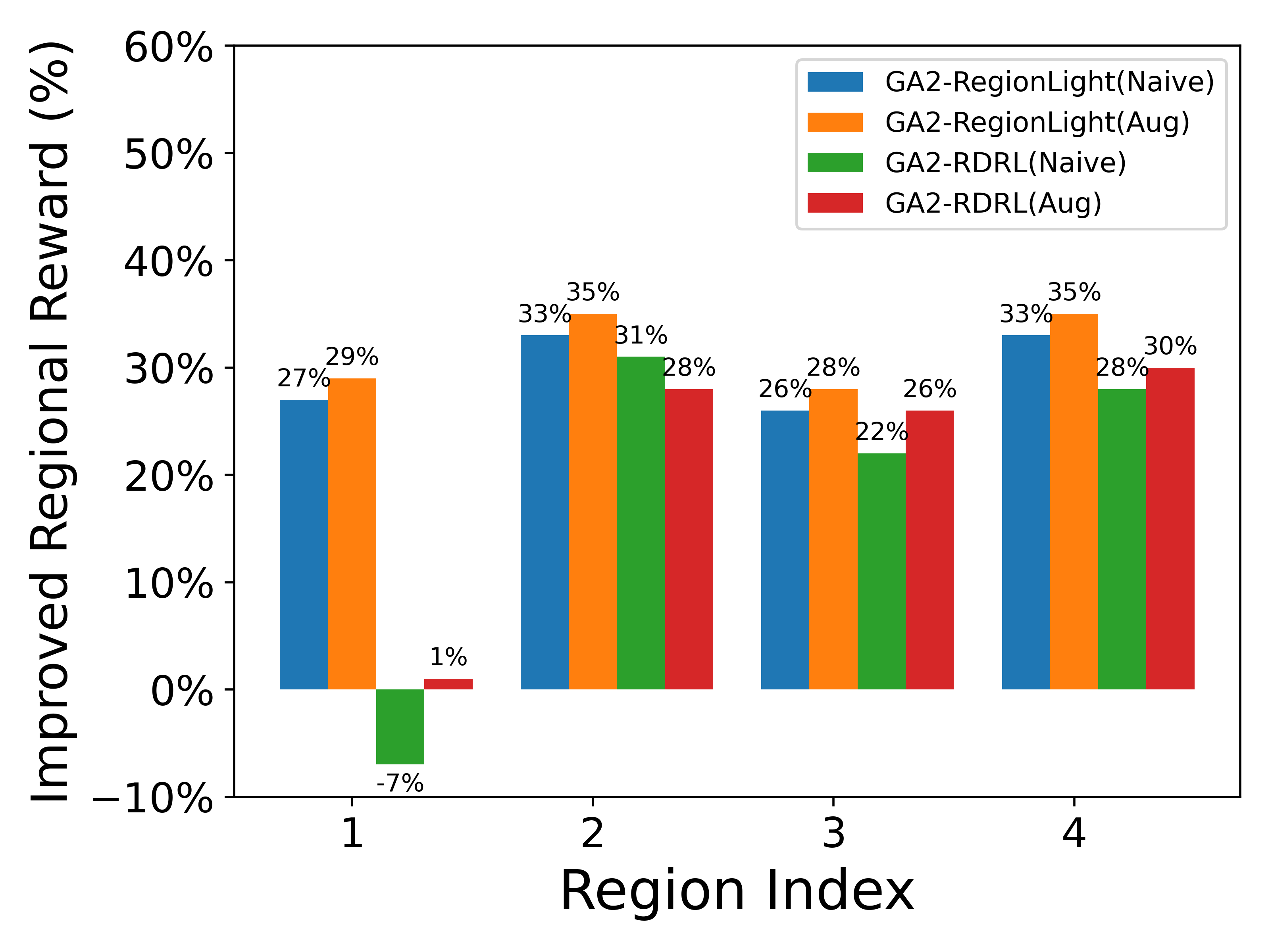}
    }
    \subfloat[New York]{
    \includegraphics[width=0.24\textwidth]{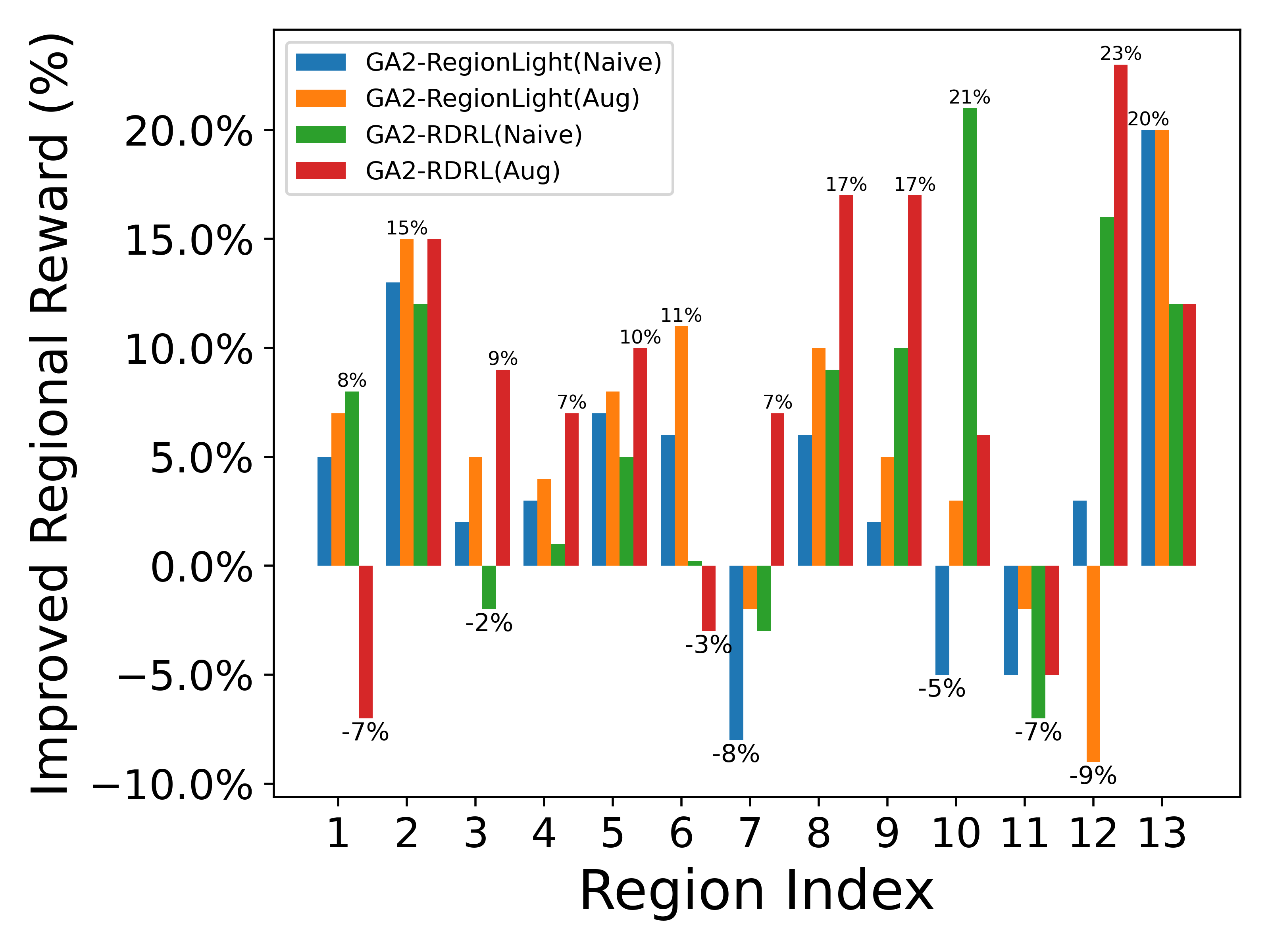}
    }
    \caption{Reward Score improvement in Regions}
    \label{fig: region improve}
\end{figure*}
The previous section evaluates the overall performance of the models using average travel time and the proposed communication modules improve the performance of existing RTSC models. This section demonstrates the improvements in reward scores of each region after the GA2 communication module is aggregated with RTSC models. The ratio of reward score improvement is depicted in Fig. \ref{fig: region improve} and the performance will be compared through each scenario. In the Hangzhou(Flat) scenario, reward scores are improved across all regions, and the improvement for RDRL is more significant than that for RegionLight. The majority of improvements are in Region 1,2,3 for GA2-RegionLight(Naive) and GA2-RegionLight(Aug). However, GA2-RDRL(Naive) and GA2-RDRL(Aug) achieve more improvements in Region 4 than RegionLight(Naive) and GA2-RegionLight(Aug). The overall improvement for GA2-RDRL(Naive) is slightly higher than that of GA2-RDRL(Aug). 
In the Hangzhou (Peak) scenario with more traffic flows, the most significant improvements in reward scores are observed in Region 2.
In the Synthetic scenario, although the performance of models is generally better than the original models, the reward scores of some regions decreased for RDRL. The reward scores of GA2-RDRL(Naive) decrease 7\% in Region 1 but increase significantly in other regions. 
In the New York scenario, the reward scores of Region 11 decrease across all models, and those of Region 7 decrease across most models except GA2-RDRL(Aug). 

Although the communication modules improve the overall performance, the benefits are not shared among all agents. This indicates that the agents face a fairness issue. Due to the natural interactions between agents, the decrease in the queue length for one intersection is probably causing an increase in the queue length of its neighboring agents. Hence, the problem is mixed with cooperative and competitive interactions. Meanwhile, our communication modules only enable agents to exchange information while the objective of agents still focuses on individual rewards. Without an explicit coordination strategy, the rewards of some agents are possible to be sacrificed.

\subsection{Impact of Cell Number and Multi-head Number}
\begin{figure*}
    \centering
    \subfloat[Hangzhou(Flat)]{
    \includegraphics[width=0.24\textwidth]{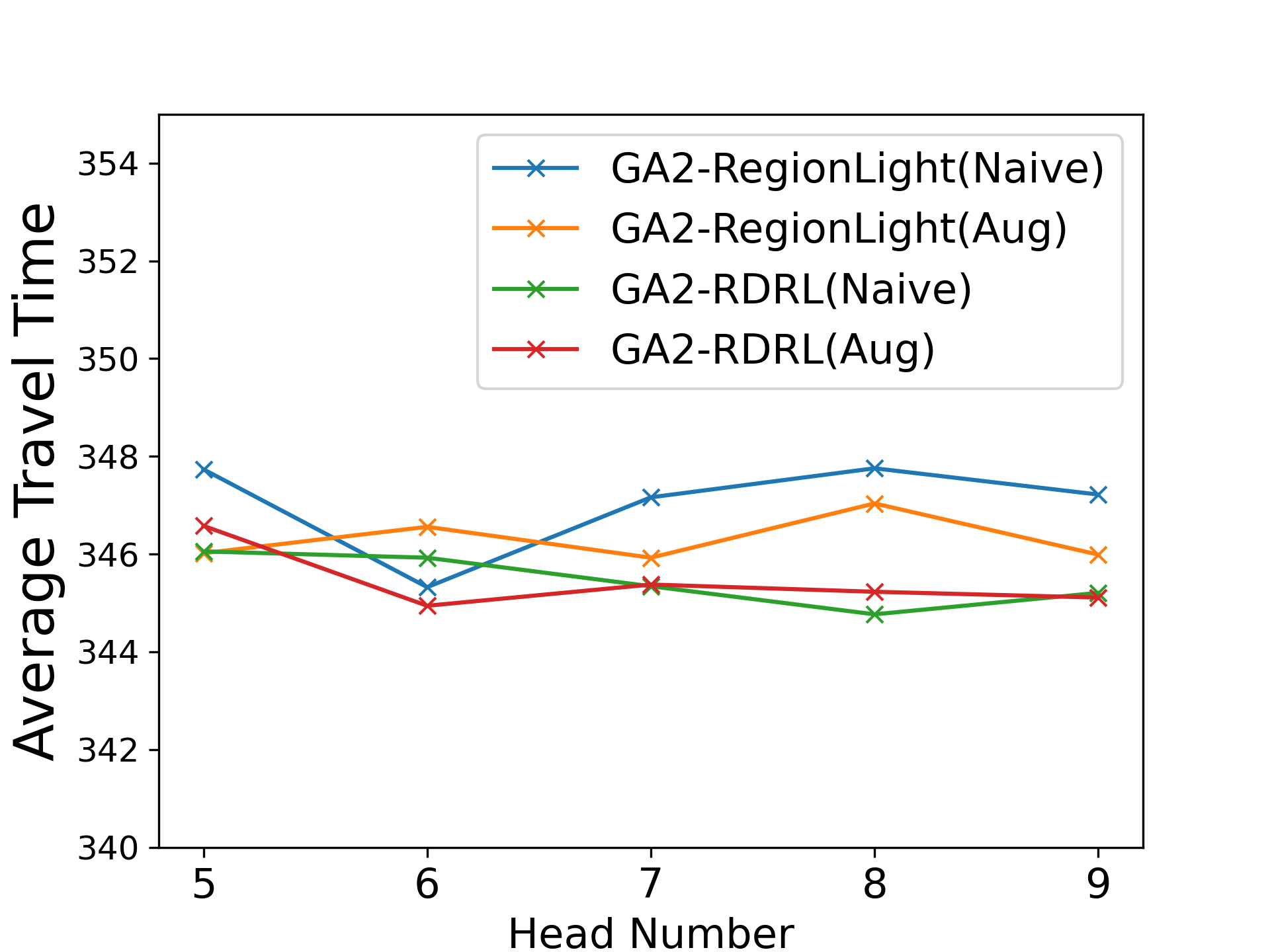}
    }
    \subfloat[Hangzhou(Peak)]{
    \includegraphics[width=0.24\textwidth]{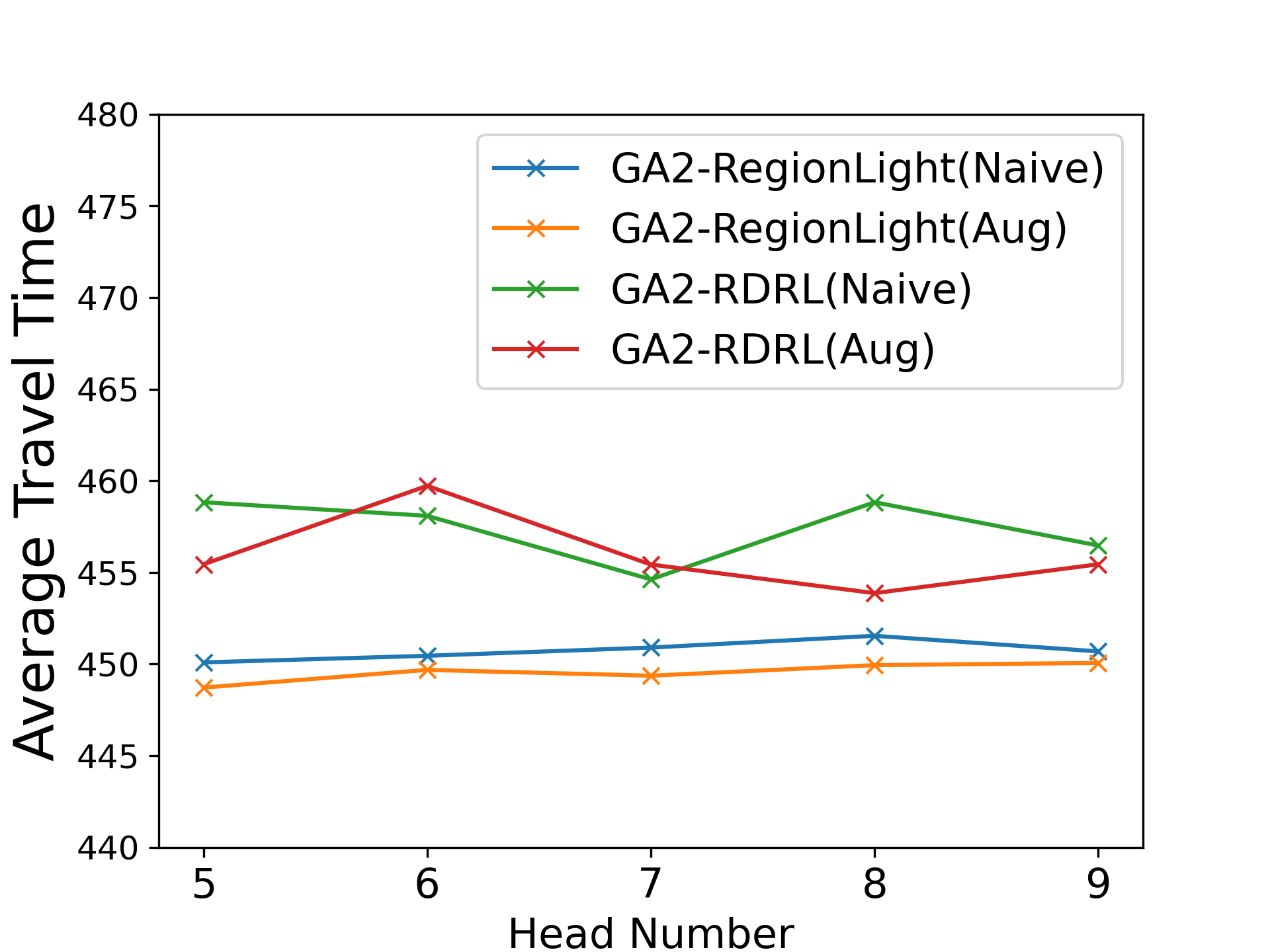}
    }
    \subfloat[Synthetic]{
    \includegraphics[width=0.24\textwidth]{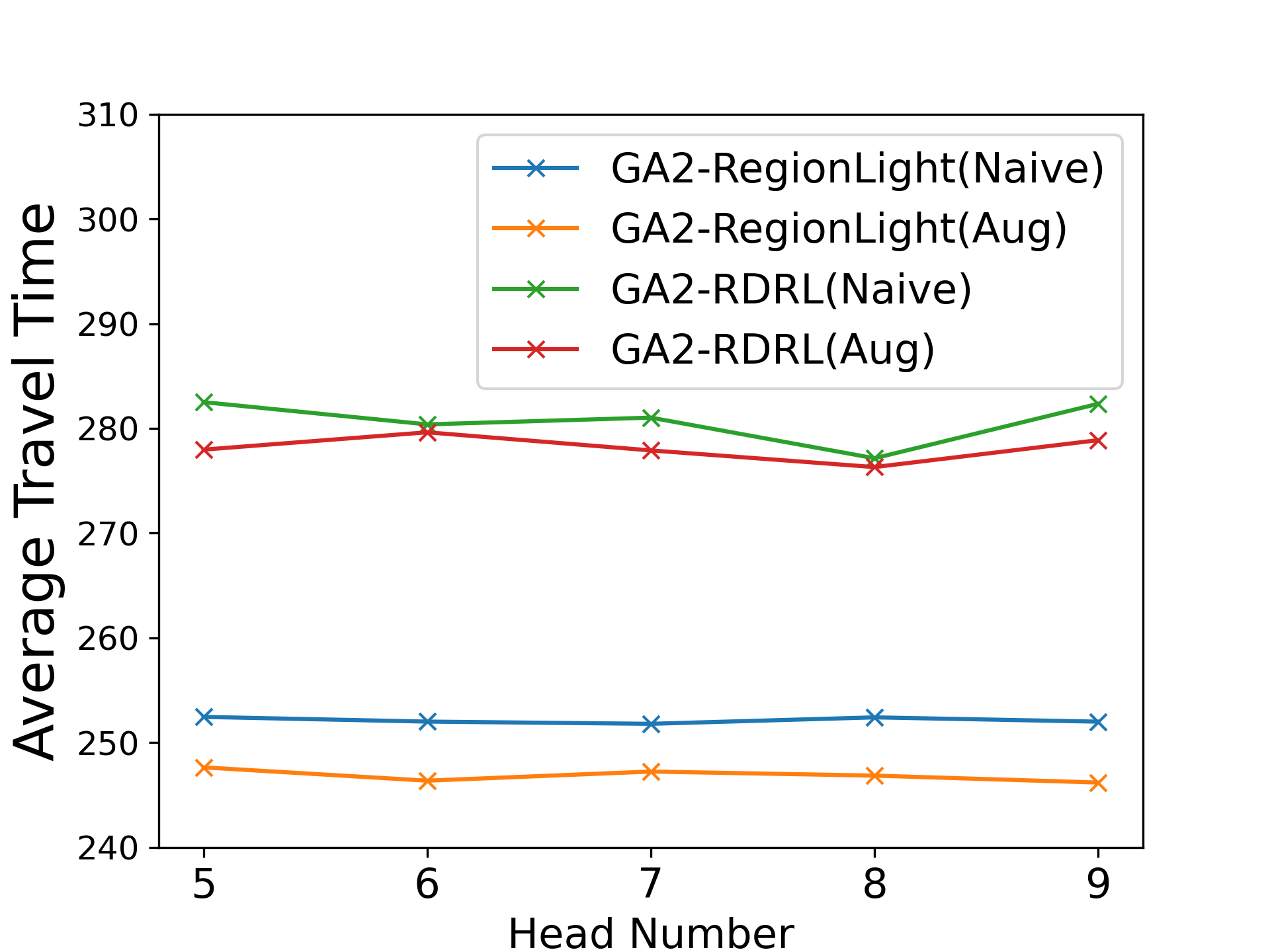}
    }
    \subfloat[New York]{
    \includegraphics[width=0.24\textwidth]{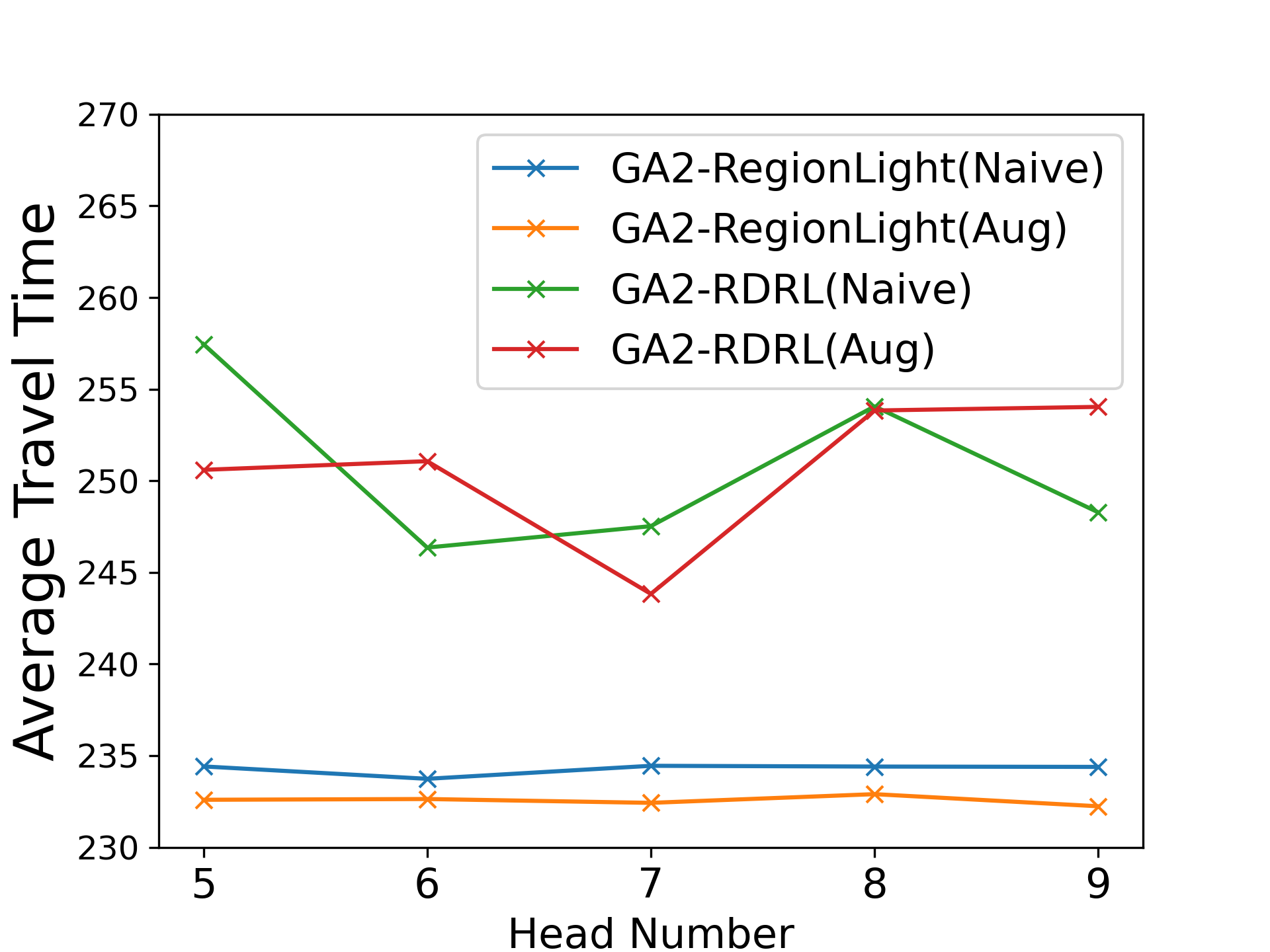}
    }
    \caption{Robustness test on different values of multi-head number}
    \label{fig: head robust}
\end{figure*}
\begin{figure*}
    \centering
    \subfloat[Hangzhou(Flat)]{
    \includegraphics[width=0.24\textwidth]{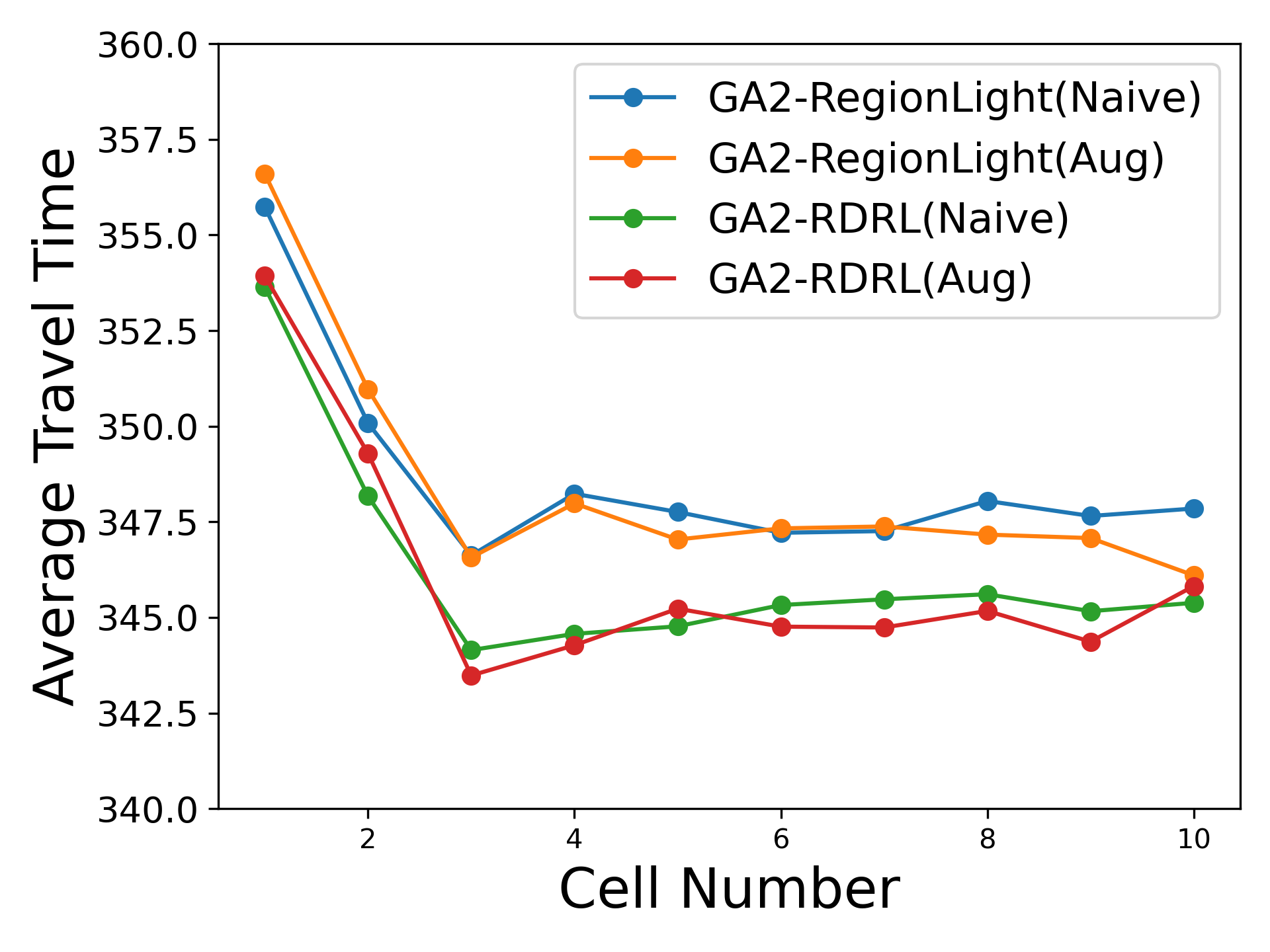}
    }
    \subfloat[Hangzhou(Peak)]{
    \includegraphics[width=0.24\textwidth]{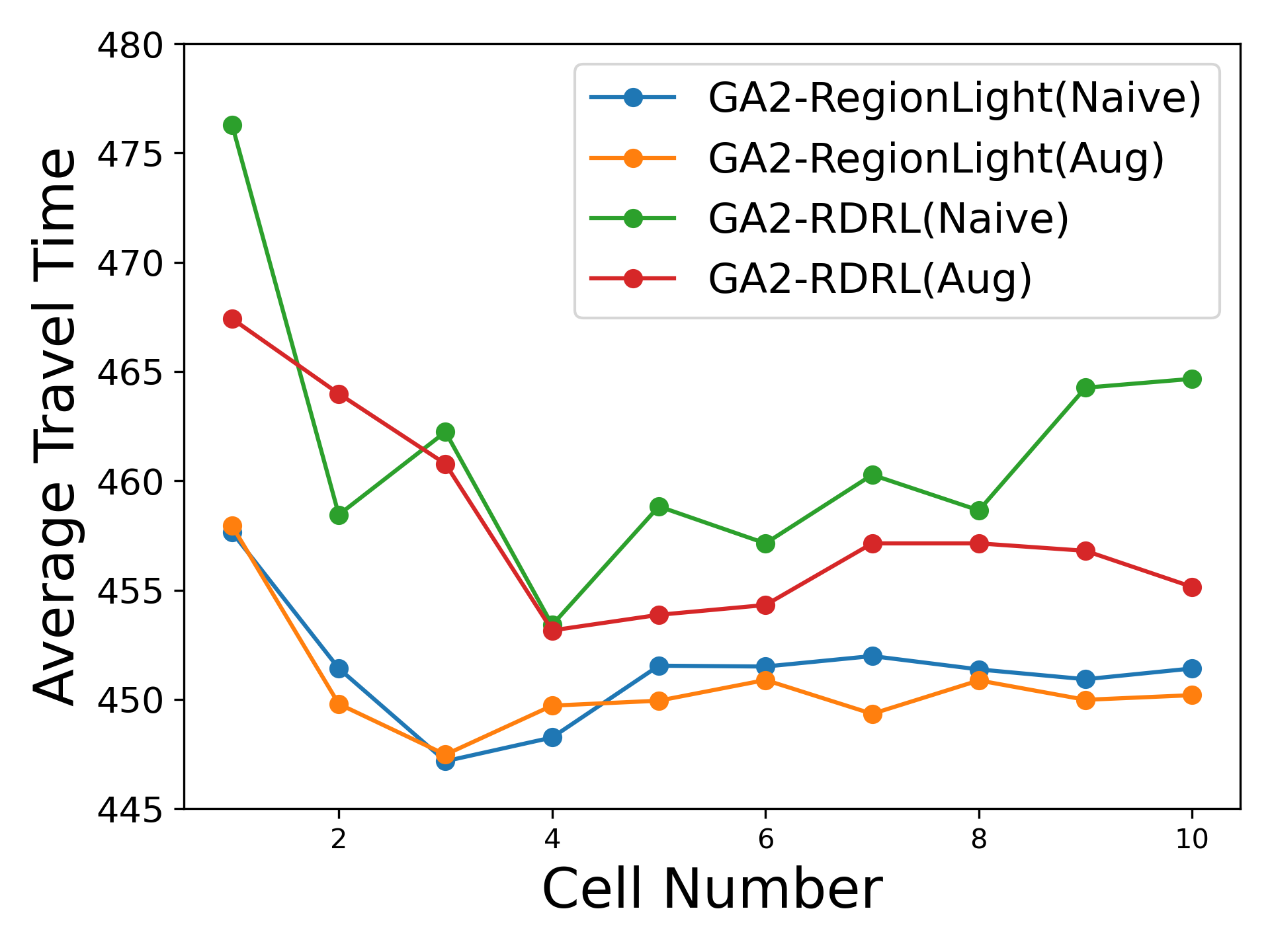}
    }
    \subfloat[Synthetic]{
    \includegraphics[width=0.24\textwidth]{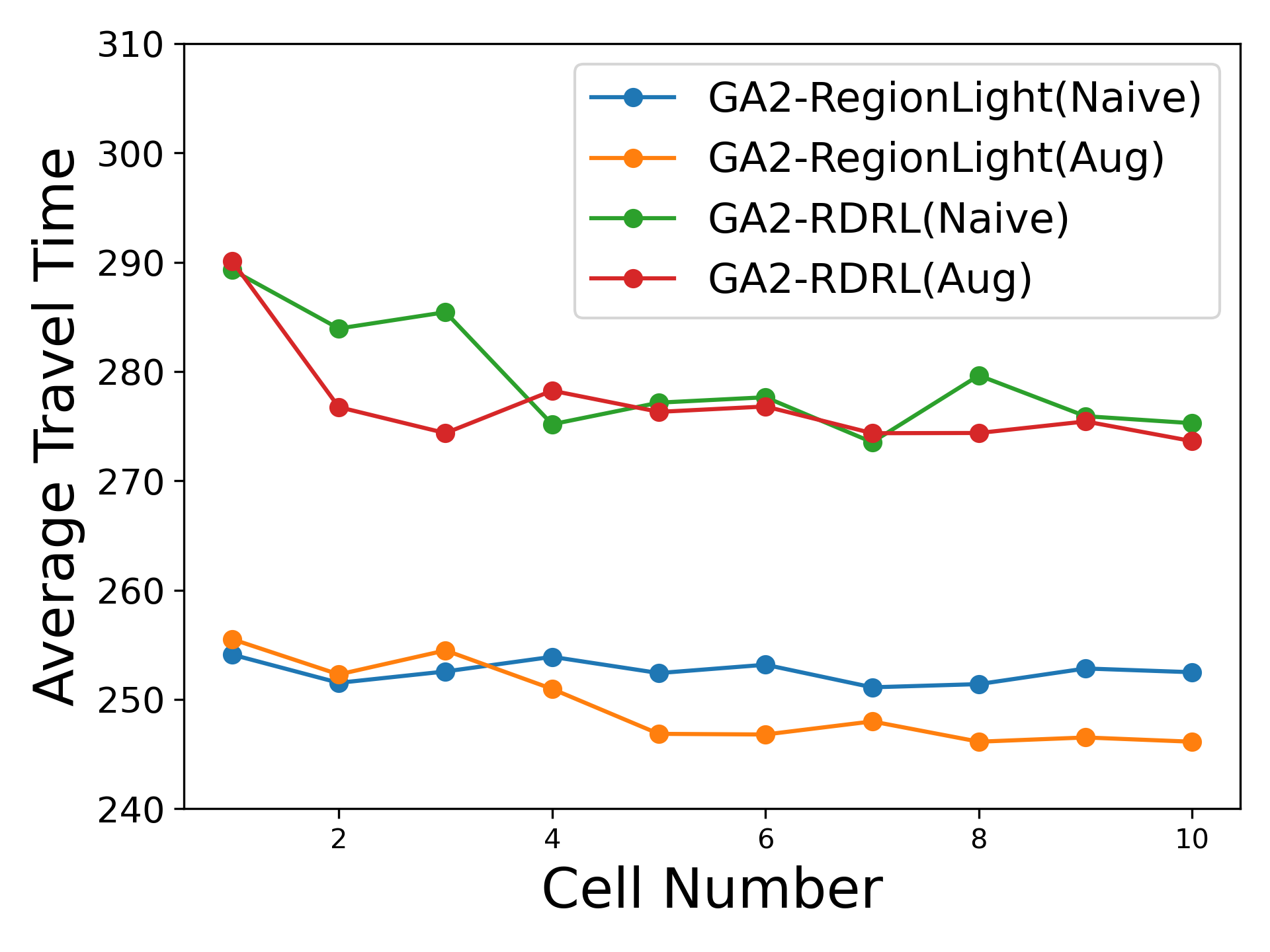}
    }
    \subfloat[New York]{
    \includegraphics[width=0.24\textwidth]{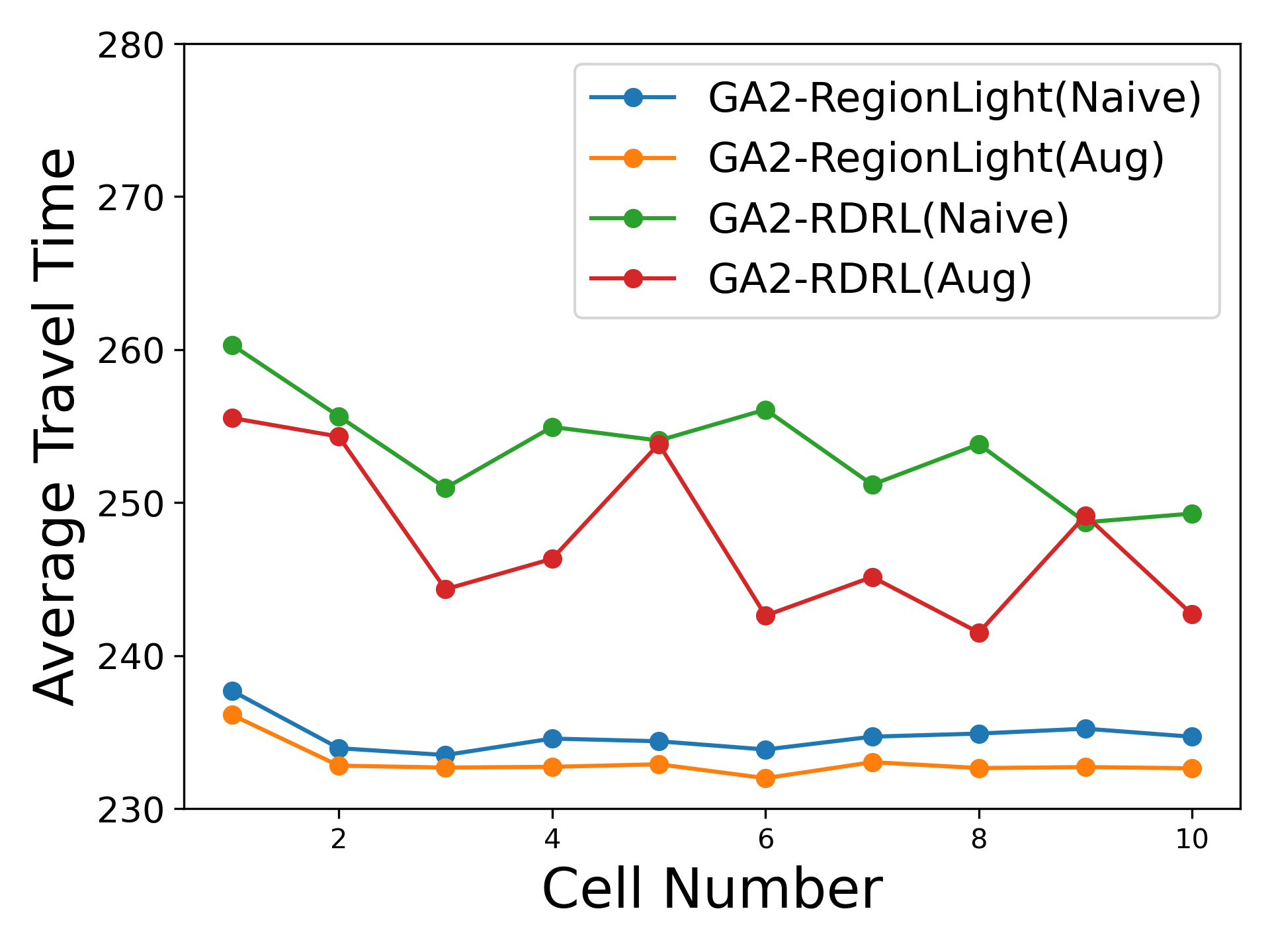}
    }
    \caption{Robustness test on different values of cell number}
    \label{fig: cell robust}
\end{figure*}

Two crucial hyperparameters are involved in the proposed communication modules. The first is the number of multi-heads in the GAT introduced in Section \ref{sec: GAT}, and the second is the number of cells introduced in Section \ref{sec: lanelevel GAT}.
In this section, we examine the impact of different numbers of headers in GAT and different numbers of cells when constructing micro traffic states. Note that the default number of cells is 5 and the default number of multi-head is 8 which are listed in Table \ref{tab: hyperparameter} and tested in previous sections.

Fig. \ref{fig: head robust} illustrates the average travel time of all scenarios when varying multi-head numbers from 5 to 9. We can observe that the performance of GA2-RegionLight(Naive) and GA2-RegionLight(Aug) remained stable under different numbers of multi-head. The performance of GA2-RegionLight(Aug) consistently outperformed GA2-RegionLight(Naive) except when the number of multi-head is 6 in Hangzhou(Flat) scenario.  However, the performance of GA2-RDRL(Naive) and GA2-RDRL(Aug) showed less stability, and more instances of tangling were observed between the curves of these two models, especially in the New York scenario. 

Fig. \ref{fig: cell robust} illustrates the average travel time of all scenarios when varying the number of cells from 1 to 10. Note that when the number of cells is 1, then the micro traffic state is equivalent to the $wave$ defined in Eq. \ref{eq: third term}. With larger cell numbers, the micro-traffic state should provide more information on different intervals on each lane. Therefore, the model has the potential to capture more useful hidden correlations. From Fig. \ref{fig: cell robust}, the performance when the cell number is 1 in all scenarios is the worst. When the number of cells increases, the performance improves. However, as the number of cells increases above 5, there is no significant improvement in all scenarios. For GA2-RegionLight models, both Naive and Augmented variants are highly stable across changes in the number of cells.
Fluctuations in performance are minimal, indicating that these models are less sensitive to hyperparameter variations. However, the performance stability of GA2-RDRL is less robust compared to GA2-RegionLight. There are noticeable oscillations in performance, particularly in the New York scenario when varying the number of cells. 
Although there are oscillations in all models across all scenarios, the performance is still better than the baselines given a not too small number of cells.

\section{Conclusion}
\label{sec: conclusion}
In this paper, to enhance the communication and cooperation between RTSC agents, we first justify that the updating equation of the RTSC process is a Markovian chain by using a system of store-and-forward queues. We then propose a novel communication module for RTSC models based on the updating equation. Our communication module leverages GAT to capture correlations between both macro and micro traffic states. The adjacent matrix of intersection is used to embed the macro traffic state. Two movement matrices are proposed to embed micro traffic states constructed by segmenting each lane into several cells. The naive Movement matrix only considers the movement at each intersection while the augmented movement matrix also considers the lane-changing behavior of vehicles on each approach.

To evaluate the proposed module, we aggregate it with two existing RTSC models. The numerical results show the aggregated models outperform baseline models and demonstrate the efficacy of our communication module in both real and synthetic scenarios. The reward scores of all regions for the aggregated models achieve improvements in the Hangzhou scenarios. Interestingly, in the synthetic and New York scenarios, the rewards of some agents are sacrificed to achieve better global performance. We also explored different settings for the number of cells and multi-heads in GAT by observing their impact on performance. However, our work still has two limitations. First, although agents share information through GA2 modules, no explicit coordination strategy is applied to guide agents' decisions. Hence, a fairness issue appears. Second is that the roadnets in our experiments are all grid roadnets. The performance of our model in more complex roadnets still remains to be explored.

In the future, we plan to investigate more aggregation techniques to capture the correlations between lane-level and intersection-level traffic states. We also plan to study the coordination strategy for RTSC models and decentralized communication strategies. Meanwhile, addressing the fairness issue among MADRL agents is also our focus. With the updating equation justified in this paper, model-based MADRL techniques are also a promising topic for further research.

\bibliographystyle{IEEEtran}
\bibliography{reference}
\end{document}